\documentclass[12pt]{iopart}
\usepackage[pdftex]{graphicx}
\usepackage{dcolumn}
\usepackage{bm}
\usepackage{epsfig}
\usepackage{dsfont}
\usepackage{bbm}
\usepackage{amsfonts}
\usepackage{amssymb}
\usepackage{amscd}
\usepackage{amstext}
\usepackage{float}
\usepackage{rotating,standalone}
\usepackage{color}
\usepackage{ulem}

\usepackage{iopams}
\usepackage{setstack}

\usepackage{tikz}

\newcommand{\expec}[1]{\langle #1 \rangle}

\newcommand{\bra}[1]{\langle\,{#1}\, |}
\newcommand{\ket}[1]{|\,{#1}\,\rangle}

\newcommand{\sub}[2]{{#1}_{ \mbox{\scriptsize #2}}}

\def\beq{\begin{eqnarray}}
\def\eeq{\end{eqnarray}}

\newcommand{\rref}[1]{Ref.~\cite{#1}}

\newcommand{\frefp}[2]{figure~\ref{#1}(#2)}

\newcommand{\cref}[1]{chapter~\ref{#1}}
\newcommand{\Cref}[1]{Chapter~\ref{#1}}
\newcommand{\aref}[1]{\ref{#1}}
\newcommand{\bref}[1]{(\ref{#1})}

\renewcommand{\emph}[1]{\textit{#1}}

\begin{document}

\title{Two-dimensional spectroscopy of Rydberg gases}
\author{K.~Mukherjee$^1$, H.~P.~Goswami$^{2,4}$, S.~Whitlock$^3$, S.~W\"uster$^1$ and A.~Eisfeld$^4$}
\address{$^1$ Department of Physics, Indian Institute of Science Education and Research, Bhopal, Madhya Pradesh 462 066, India}
\address{$^2$ Department of Chemistry, Gauhati University, Jalukbari, Guwahati, 781014, Assam, India}
\address{$^3$ ISIS (UMR 7006), University of Strasbourg and CNRS, 67000 Strasbourg, France}
\address{$^4$ Max Planck Institute for the Physics of Complex Systems, N\"othnitzer Strasse 38, 01187 Dresden, Germany}
\ead{sebastian@iiserb.ac.in, eisfeld@pks.mpg.de}
\begin{abstract}
Two-dimensional (2D) spectroscopy uses multiple electromagnetic pulses to infer the properties of a complex system. A paradigmatic class of target systems are molecular aggregates, for which one can obtain information on the eigenstates, various types of static and dynamic disorder and on relaxation processes.  However, two-dimensional spectra can be difficult to interpret without precise knowledge of how the signal components relate to microscopic Hamiltonian parameters and system-bath interactions. Here we show that two-dimensional spectroscopy can be mapped in the microwave domain to highly controllable Rydberg quantum simulators. 
By porting 2D spectroscopy to Rydberg atoms, we firstly open the possibility of its experimental quantum simulation, in a case where parameters and interactions are very well known. Secondly, the technique may provide additional handles for experimental access to coherences between system states and the ability to discriminate different types of decoherence mechanisms in Rydberg gases.
 We investigate the requirements for a specific implementation utilizing multiple phase coherent microwave pulses and a phase cycling technique to isolate signal components.
 \end{abstract}

\maketitle

\section{Introduction}
\label{sec_intro}

Assemblies of molecules can exchange electronic excitation energy via long-range interaction of transition dipoles \cite{book:maykuehn}.
This process leads to eigenstates that are coherently delocalised over several molecules, where the details depend on the molecular arrangement and the interaction with internal vibrations and the environment. Prominent examples are J- and H-aggregates of organic dyes \cite{saikin:excitonreview} or the light harvesting systems in photosynthesis \cite{grondelle:review}.

To understand the resultant eigenstates and dynamics in molecular systems, multi-dimensional optical spectroscopy has become an indispensable tool \cite{mukamel:book,mukamel2000m,TeDyMa06_194303_,NeMiMa10_094505_,PaMaAs15_212442_,MiPrMa13_6007_,BrHiKoe17_1700236_}. 
A sequence of ultrashort laser pulses with well defined relative phases results in a signal that is measured as a function of the time-delays between the pulses.  This signal may for example be the population of system states or electromagnetic radiation detected in a specific direction. 
Typically, a Fourier-transform with respect to some (or all) time delays yields a spectrum in the frequency domain.
The most common technique, referred to as two-dimensional (2D) spectroscopy, employs four pulses and thus three time delays. 
Usually, the Fourier transformation is made with respect to the first time interval $\tau$ (between the first and second pulse) and the last time interval $t'$ (between the third and fourth pulse). 
One obtains a separate 2D spectrum for each time delay between the second and third pulse, called the waiting time $T$.
This sequence of spectra holds a multitude of information about the dynamics of the system, its coherences and relaxation pathways as well as different line broadening mechanisms, see e.g.~Ref.~\cite{BrHiKoe17_1700236_}.

However, these spectra are difficult to interpret and one easily can draw flawed conclusions based on incomplete or inappropriate models of the system under study.
Numerical calculations should on the one hand reproduce the experimental data, on the other hand they should provide insight into how the signal depends on the microscopic properties of the system, e.g.~by systematically changing the model parameters. Unfortunately due to the theoretical complexity  this is not possible for many systems of interest without making severe approximations, especially regarding the treatment of internal vibrations and the coupling to an environment. Although sophisticated numerical methods have been developed to treat the dynamics of molecular systems (e.g.\ Refs.~\cite{KaTa04_260_,RiRoSt11_113034_,Suess_HOPS_PRL,KrKrRo11_2166_}) in practice one can still handle only a small regime of parameters and model Hamiltonians.

An alternative to numerical simulations might be offered by quantum simulators, well controllable physical systems sharing a common Hamiltonian with the target system that can be used to study quantum dynamics in regimes where numerical simulations fail.
There have been several proposals of quantum simulators for assemblies of interacting molecules \cite{schoenleber:immag,MoReEi12_105013_,MoHuKr16_44_,plodzien_polaronbio_SciRep,herrera2011tunable,potocnik_lightharvest_SC,PhysRevLett.123.100504,PhysRevX.8.011038}, which could also benefit from sophisticated probing techniques such as multidimensional spectroscopy. Enabling the latter on these platforms, would finally facilitate direct comparisons with real molecular systems. In the present work we propose a quantum simulator based on dipole-coupled assemblies of Rydberg atoms that closely mimics the physics of small molecular aggregates and is compatible with multidimensional spectroscopy techniques.

In the case of molecular assemblies, the transport of electronic excitation energy proceeds on ultrafast timescales by optical dipole couplings between molecular subunits on sub-wavelength distance scales~\cite{AmVaGr00__,BrHiKoe17_1700236_,saikin:excitonreview}. In a Rydberg quantum simulator, analogous dipole-dipole interactions in the microwave regime allow the migration of Rydberg excitations through an assembly of atoms on microsecond and micrometer scales~\cite{barredo:trimeragg}.
In contrast to the molecular case, the monomers (atoms) in the quantum simulator can be positioned nearly at will \cite{nogrette:hologarrays,wang_tweezerarray}, their transition energies and interactions tuned through choice of Rydberg quantum state \cite{book:gallagher} and Markovian \cite{schempp:spintransport,schoenleber:immag} or non-Markovian \cite{genkin:markovswitch,schoenleber:thermal,BeEi18_205003_} environments can be engineered to simulate coupling of an excitation to vibrations. Because transitions between the relevant Rydberg states correspond to microwave frequencies, it is not a-priori clear, whether parameters such as interaction strength, lifetime, dephasing rates and interrogation pulse durations and spacings scale well from the molecular setting to the Rydberg setting such that the technique remains feasible.
 We show in the following that it is indeed possible to find a suitable parameter regime.

In this way we establish a clean platform for testing basic features of multi-dimensional spectroscopy, in which a large variety of disorder and decoherence sources can be engineered  \cite{schoenleber:immag,schempp:spintransport,genkin:markovswitch,schoenleber:thermal,BeEi18_205003_}. 
 Our main focus in this article is to adapt 2D spectroscopy to ultracold Rydberg systems in order to understand and simulate the corresponding situation in optical spectroscopy of molecular systems. In addition 2D spectroscopy would provide an additional diagnostics tool that may complement the already extensive arsenal available to ultracold Rydberg physics \cite{lim:rydbergreview,loew:rydguide:jpbreview}.

While we focus on the specific case of Rydberg atoms, microwave domain 2D spectroscopy can more generally be useful whenever quantum dynamics involves energy differences in the microwave realm, such as in quantum dots \cite{oosterkamp1998microwave}, Nitrogen-vacancy-centers or superconducting circuit arrays \cite{potocnik_lightharvest_SC,MoReEi12_105013_,MoHuKr16_44_}, and much of this section should apply to these systems too. The technique is already being used, for example, in rotational spectroscopy of poly-atomic molecules \cite{Wilcox_2Dchirps_JCA} and collisional population transfer \cite{vogelsanger_2D}.

 To isolate signals emerging from a specific non-linear process, 2D spectroscopy of solid state systems or large molecular samples often involves a so-called phase matching approach  \cite{book:boyd_nonlinopt,mukamel:book,tan:2dcycling}. Since it would be challenging to detect such micro-wave signals in the typically very dilute Rydberg assemblies, we propose to record signals based on Rydberg state populations. We thus explore what is called action detection based 2D spectroscopy with an 
incoherent action detection method. An important consequence of this, is the unavailability of phase matching in this type of microwave spectroscopy. We propose to use phase cycling \cite{dorineORIG,TiKeSu03_1553_,tan:2dcycling}  with a fast measurement of the final excited state populations.

\begin{figure}[htb]
\centering
	\includegraphics[width=0.99\columnwidth]{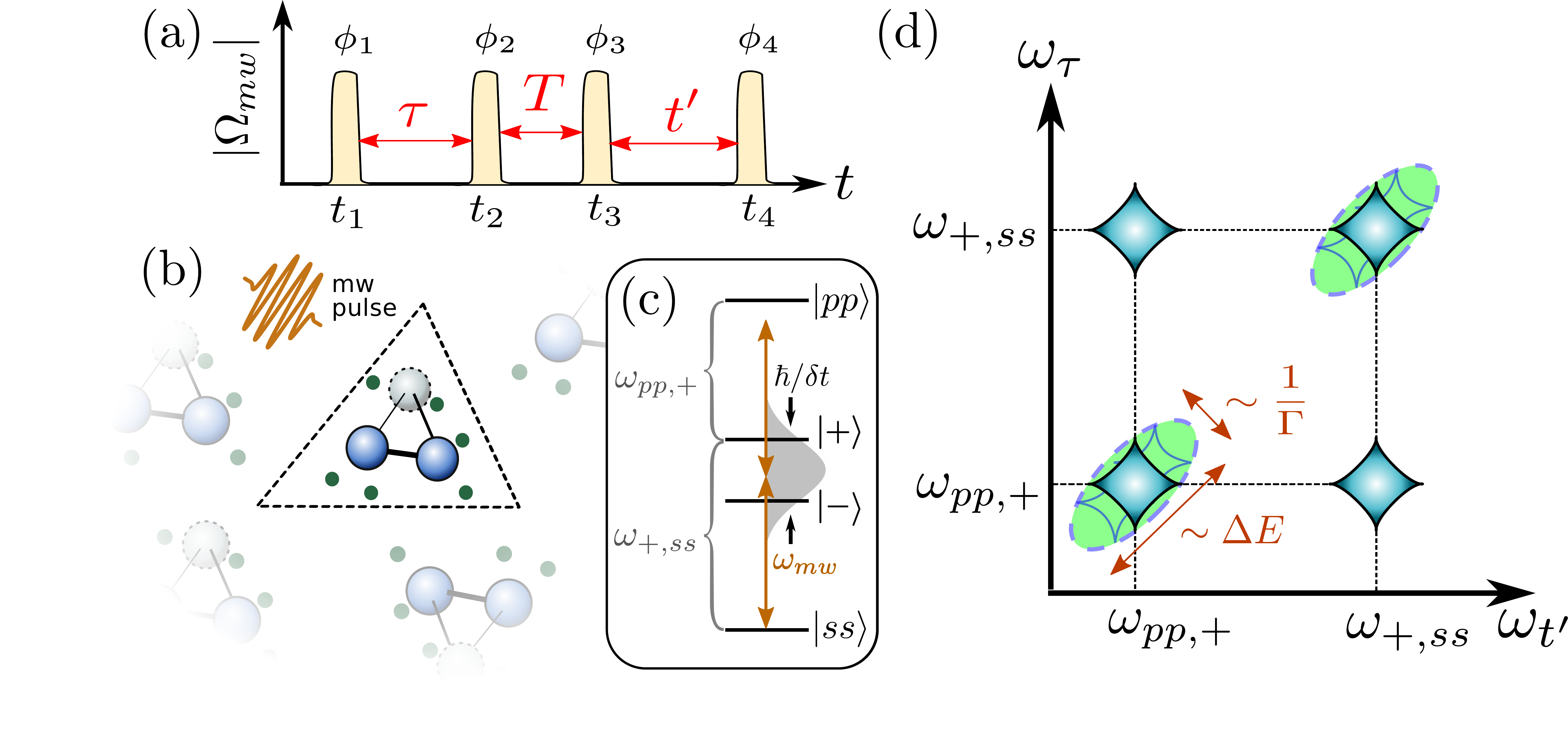}
\caption{(a) Schematic of a 4-pulse microwave sequence, with variables discussed in the text. (b) Sketch of Rydberg-dimer (blue) or trimer (blue+grey), to be interrogated by the microwave pulses, while green ultracold ground state atoms provide tuneable disorder and decoherence. (c) For the specific example of a Rydberg dimer, four two-body states play a role. Variables are discussed in the text. (d) The corresponding energy differences show up in a 2D spectrum as diagonal peaks, with off-diagonal peaks indicating coherences between them. Inhomogeneous broadening ($\sim \Delta E$) and homogeneous broadening ($\sim 1/\Gamma$) can be distinguished as indicated, since they affect different frequency axes in the 2D spectrum.
 \label{sketch}}
\end{figure}

This article is organized as follows: In section \ref{microwavespec}, we present the general method of 2D microwave spectroscopy involving a series of short microwave pulses interacting with a set of coupled two-level systems. We discuss the pulse sequence, spectroscopic signal and phase cycling method for a measurement of the nonlinear response of the system briefly in that section, with further details in
 \aref{app_perttheory}, \ref{app_phasecycling} and \ref{photon_echo}. In section \ref{rydaggspec} we discuss how to port microwave 2D spectroscopy to the field of ultracold Rydberg gases. We show how decoherence in the system can be experimentally controlled in these systems and derive effective equations for its description in \ref{app_LeffHeff}. In section \ref{results} we present proof-of-principle numerical calculations of 2D spectra for simple models like Rydberg dimer and Rydberg trimer aggregates, which are embedded within an environment of perturbing atoms giving rise to both homogenous and inhomogeneous broadening mechanisms. We show that experimental constraints arising from the total spectrum acquisition time might require special consideration, possibly necessitating parallel interrogation of many identical Rydberg aggregates. Finally, we discuss future prospects for Rydberg quantum simulators and the experimental simulation of photosynthetic complexes.

\section{Generic aspects of microwave 2D spectroscopy}
\label{microwavespec}

In this section, we briefly outline the basic scenario that we propose and provide some general principles of microwave 2D spectroscopy.
We keep this discussion quite general.
Details specific to Rydberg quantum simulators are discussed in the subsequent \sref{rydaggspec}. 

\subsection{The system and its environment}
\label{microwave}

We consider a collection of $N$ (long-range) interacting two-level systems.
We are in particular interested in a situation where the energy scale of the interaction is small compared to the transition energies between the two states of the two-level systems. 
An incoming electromagnetic field can drive transitions between these states.
For weak fields it is convenient to classify the eigenstates of such a system according to the number of 'excitations', i.e, the number of excited two level systems.
We will be mainly be interested in the subspaces with zero, one or two excitations, which are usually denoted as the ground state, the single exciton manifold and the two-exciton manifold, respectively.
Typically, there are two effects that complicate the interpretation of  measurements on this system:
1.) The measurement is performed on an ensemble of non-identical systems (static disorder). Such an ensemble emerges for example via imperfect preparation, sample inhomogeneity or fluctuations that are slow compared to the interrogation time scale.
2) Coupling to additional degrees of freedom that leads to dephasing and  relaxation.
Both effects usually result in broadening of absorption lineshapes.
One strength of 2D spectroscopy is the ability to disentangle the contributions of these two effects.

Because of the coupling to the environment, the dynamics of the system is typically described by a time-evolution equation for the reduced density matrix of the system $\rho_\mathrm{S}(t)$, which can be formally written as 
\begin{eqnarray}
\rho_\mathrm{s}(t)=\mathcal{U}(t) [\rho_\mathrm{S}(0)],
\label{generic_evolution_equation}
\end{eqnarray}
 where $\mathcal{U}(t)$ is a time-evolution super operator.

\subsection{Pulse sequence}

This system is irradiated by a sequence of electromagnetic pulses, which are characterized by a time-dependent electric field $\mathbf{E}(t)$. The electric field $\mathbf{E}(t)$ is assumed to be given by a train of four short pulses
\begin{eqnarray}
\mathbf{E}(t) &= \mathbf{E}_0 \sum_{j=1}^4 A(t  - t_j)\cos{(\omega_0 t - \varphi_j)},
\label{Eoft}
\end{eqnarray}
where the $A(t)$ are the (slowly varying) pulse-shapes, roughly of duration $\delta t$, and $\varphi_j$ control the relative phases of pulses, as sketched in \fref{sketch}(a), and $\mathbf{E}_0$ their amplitude and polarisation vector. We define time delays as in \fref{sketch}, such that $\tau=t_2-t_1$, $T=t_3-t_2$, $t'=t_4-t_3$.

The carrier frequency $\omega_0$ is chosen to be close to the relevant transition energies $h_n$ of the two-level systems. For a system with spatial extent much smaller than the wavelength, the interaction of the system with impinging radiation in the dipole approximation takes the form
\begin{eqnarray}
\sub{\hat{H}}{mw}&=\sum_n \hat{\mu}^{(n)} E(t),
\label{Hint_generic}
\end{eqnarray}
where $\hat{\mu}^{(n)}$ is the relevant dipole operator of object $n$ and we have 
assumed a linearly polarized field aligned parallel with the transition dipole moments, using $E(t)=|\mathbf{E}(t)|$. Furthermore, we assume that the wavelength is large compared to the spatial extent of the system. All discussed assumptions will be fulfilled for the Rydberg setup presented in the following sections. The matrix elements of the total system dipole operator $\hat\mu\equiv \sum_n \hat{\mu}^{(n)}$ are used to determine whether or not transitions between eigenstates are allowed. 
 
The duration of each pulse $\delta t$ and the delays $\tau$, $t'$ are  dictated by the details of the spectrum of $\sub{\hat{H}}{sys}$ to be interrogated, since pulse durations and delays have to be chosen to resolve the interactions of interest. 
We thus require 
\begin{eqnarray}
\delta t \lesssim \frac{\hbar}{\mbox{max}|{\cal E}_\ell-{\cal E}_{\ell'}|},
\label{pulsewidths}
\end{eqnarray}
where $\cal E_\ell$ and $\cal E_{\ell'}$ belong to the same exciton manifold, 
so that microwave pulses are spectrally wide enough to reach all transitions in the system, see \frefp{sketch}{c}, and $\pi/\tau$, $\pi/t' \lesssim \sigma_\omega$, where $\sigma_\omega$ is the linewidth of a certain transition, in order for the spectrum to resolve details of resonance peaks.

\subsection{Spectroscopic signal and phase-cycling}
\label{signal_and_cycling}

In \textit{optical} multi-dimensional spectroscopy, the signal of interest generated by the evolution of $\hat{\rho}_s(t)$ can frequently be isolated by choice of a specific geometry of incoming beam directions and outgoing signal direction, exploiting the phase-matching enforced by the propagation of fields through the interrogated medium. 

For many of the quantum systems amenable to microwave interrogation that we listed in \sref{sec_intro}, there will be no significant propagation through a medium and thus phase-matching is not readily available. On the other hand, these systems often allow additional access to some observables within the system, for example the determination of the total spin projection.
Let in the following $\hat{F}$ denote any observable, measured immediately after the last pulse, at time $\sub{t}{end}$. 
The signal depends crucially on the time delays $\tau$, $T$, $t'$ and the three relative phases between subsequent pulses.
We write
\begin{eqnarray}
 \label{signal}
S_\xi(\tau, T, t') = \expec{\hat{F}(\sub{t}{end})}
\end{eqnarray}
where $\xi$ is an abbreviation for the triplet of relative phases.
 In recent years the so-called 'action detection' method for optical/visible 2D spectroscopy have become popular where the operator $\hat{F}$ can represent fluorescence intensity \cite{TeDyMa06_194303_,TeLoMa07_214307_,BrMuSt15_23877_}, photoelectrons \cite{BrEiBa19_-_,aeschlimann2011coherent}, photocurrents \cite{nardin2013multidimensional,karki2014coherent}
and photo-ions \cite{roeding2018coherent}.
For a recent work discussing differences between phase sensitive detection and action-based detection methods see Ref.~\cite{kuehn_interpreting_JPCL}.
In the Rydberg case discussed in the next section the signal is chosen to be the excited state populations directly after the last pulse, which one can for example measure via field-ionization.

 The signal \bref{signal} depends on the strength of the electromagnetic radiation, and for weak fields an expansion with respect to the electric field strength can be used. The signal is then typically dominated by the lower order contributions, while in 2D spectroscopy one is interested in specific higher order contributions.
One particular technique to extract these is  {\it phase cycling}  \cite{dorineORIG,TiKeSu03_1553_,tan:2dcycling}.
In this approach the signal $S_\xi(\tau, T, t')$ is measured for several  specific choices of the relative phases $\xi$.
By choosing suitable phase combinations one can isolate a specific signal of interest by a summation of individual signals for the pulses 
\begin{eqnarray}
 \label{cycledsignal}
S(\tau,T,t')&=\sum_\xi  c_\xi S_\xi(\tau,T,t'), 
\end{eqnarray}
where the $c_\xi$ coefficients are determined by the choices of $\xi$.
In optical 2D spectroscopy there has been much interest in the so-called  rephasing signal \cite{mukamel:book}.
Throughout this article, we will use  the rephasing signal as our example to illustrate microwave Rydberg 2D spectroscopy , which is formally defined by a set of Liouville pathways \cite{mukamel:book,kuehn_interpreting_JPCL} discussed in \ref{app_perttheory}. 
We briefly introduce phase-cycling in \ref{app_phasecycling}, and discuss how the phases and the coefficients $c_\xi$ have to be chosen to isolate it.
Other signal classes at the same order of perturbation theory are the 'non-rephasing' and the 'double quantum coherence' signals, see for example \rref{kuehn_interpreting_JPCL} and the supplemental material.

As the final step in 2D spectroscopy, we Fourier transform the time-domain signal with respect to $\tau$ and $t'$
to reach the frequency domain 2D spectrum 
\begin{eqnarray}
\tilde{S}(\omega_{\tau},T,\omega_{t'})= \int_0^\infty\int_0^\infty S(\tau,T,t') e^{-i \omega_{\tau} \tau} e^{i \omega_{t'} t'} d\tau dt'.
\end{eqnarray}
 This signal is a complex quantity, and can hence be analysed in terms of real and imaginary parts or absolute value and phase. 
We show here absolute values and phase, and in the supplemental material additionally present real and imaginary parts of all spectra shown in this article.

\section{Rydberg aggregate 2D spectroscopy}
\label{rydaggspec}

We now move to the specific application of the 2D spectroscopy protocol to Rydberg aggregates.

\subsection{Rydberg Aggregate}
\label{ryd_aggregate}

As a Rydberg aggregate \cite{wuester:review} we consider a collection of dipole-dipole coupled atoms, where each atom can be in either of two electronic states $\ket{s}$ or $\ket{p}$, with energies ${\cal E}_s$ and ${\cal E}_p$, respectively.
 We choose $\ket{s}=\ket{\nu=43,l=0,m=0}$ and $\ket{p}=\ket{\nu=43,l=1,m=0}$, where $\nu$ is the principal quantum number and $l$ the angular momentum. Many other pairs of dipole coupled Rydberg states would be equally useful. The Hamiltonian for the aggregate including dipole-dipole interactions is then ($\hbar=1$):
\begin{eqnarray}
\sub{\hat{H}}{agg}&= \sum_{nm} W_{nm}  \hat{\sigma}^{(n)}_{ps}  \hat{\sigma}^{(m)}_{sp}  +\sum_n (\epsilon_{s} \hat{\sigma}^{(n)}_{ss} +\epsilon_{p} \hat{\sigma}^{(n)}_{pp}),
\label{Hagg}
\end{eqnarray}
where $\hat{\sigma}^{(n)}_{ab} = [\ket{a}\bra{b}]_{n}$ is a transition operator on atom $n$.
In general, dipole dipole interactions would allow transition to additional states with $m\neq0$ \cite{moebius:cradle}, which can be suppressed by Zeeman-shifting them out of resonance with a magnetic field, see e.g.~\cite{Ravets:angdep:PhysRevA.92.020701,leonhardt:orthogonal}. Then, 
 \begin{equation}
 W_{nm}=W(\mathbf{r}_n,\mathbf{r}_m) =C_3(1-3\cos^2 \theta) /|\mathbf{r}_n - \mathbf{r}_m|^3
 \label{dipole_dipole_int}
 \end{equation}
 describes the strength of dipole-dipole coupling between atom pairs located at $\mathbf{r}_n,\mathbf{r}_m$, with angle $\theta$ between the direction of transition dipoles and the inter-atomic axis. The transition dipole  between the two states defined above points along the quantisation axis, given by the direction of the applied static magnetic field. 

Let us now consider additional microwave irradiation of that aggregate, assuming the microwave field to be linearly polarized along the quantisation axis.
The central frequency $\omega_0$ of the microwave pulses is chosen to be roughly resonant with the transition energy between these two states $\hbar \omega_0=\hbar\omega_{sp}=\epsilon_p-\epsilon_s$.  All frequency axes for spectra shown later in this article are relative to $\omega_0$. For the example above  $\omega_0/(2\pi)=49$ GHz. 
 Using $\hat{\mu}^{(n)}= \mu_0( \hat{\sigma}_\mathrm{sp}^{(n)} + \hat{\sigma}_\mathrm{ps}^{(n)})/2$  in \bref{Hint_generic}, the Hamiltonian for light-matter interaction is, after conversion to a rotating frame at the field-frequency and performing the rotating wave approximation:
\begin{eqnarray}
\sub{\hat{H}}{mw}&= \sum_n [\mu_0 \tilde{E}(t) \hat{\sigma}^{(n)}_{sp} + \mbox{h.c.} - \sub{\Delta}{mw}  \hat{\sigma}^{(n)}_{pp}].
\label{Hmw}
\end{eqnarray}
 Here, $\mu_0$ is the transition dipole moment between the two states and $ \tilde{E}(t)$ is the complex electric field envelope of the microwave pulses,
\begin{eqnarray}
\tilde{E}(t) &= E_0 \sum_{j=1}^4 A(t  - t_j)\exp{[-i\varphi_j]}.
\label{Eoft_complex}
\end{eqnarray}
 By $ \sub{\Delta}{mw} =\sub{\omega}{mw}-\omega_0$ we denote the microwave detuning. 

Spatially, all atoms are assumed to be trapped in the quantum ground-state of very tight optical traps. Thus interaction parameters in \bref{Hagg} are effectively spatially averaged over the 
ground-state wavefunctions. Recent experiments can position these optical traps almost at will, using digital micromirror devices or spatial light modulators \cite{nogrette:hologarrays,wang_tweezerarray}.

We assume the Rydberg aggregate is initialized in the state $\hat{\rho}_{\boldsymbol{s}} = \ket{\boldsymbol{s}}\bra{\boldsymbol{s}}$, where $\ket{\boldsymbol{s}}=\ket{ss..s}$, with all aggregate atoms in $\ket{s}$. It then interacts with the microwave-pulse train \bref{Eoft}, which will drive transitions to states which contain a non-zero $p$ state population.

Most ultracold Rydberg experiments can routinely measure the total number of atoms in a specific Rydberg state quite accurately. We thus chose $\hat F=\sum_n [|p\rangle\langle p|]_n$ as measurement operator, which counts the number of $p$-excitations generated in the system by the microwave pulse train.

\subsection{Controllable environment for decoherence}
\label{environment}

To assess the effects of decoherence we embed the Rydberg aggregates in a cold atom environment that allows tailoring a variety of different decoherence mechanisms \cite{schoenleber:immag,schempp:spintransport,genkin:markovswitch,schoenleber:thermal}.

\begin{figure}[htb]
\centering
\includegraphics[width=0.9\columnwidth]{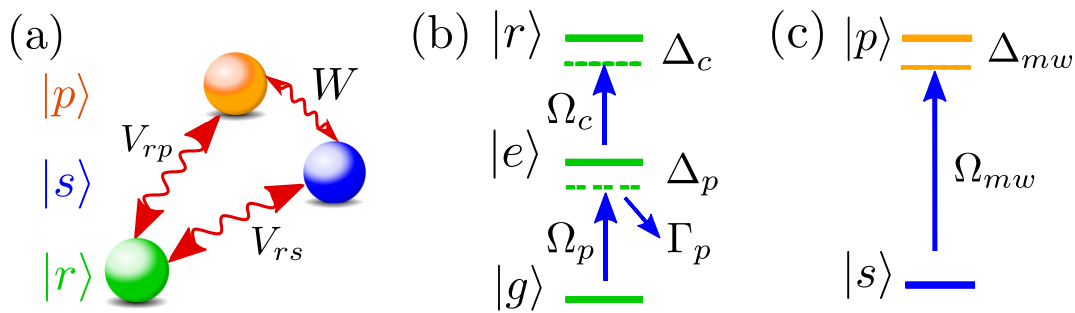}
\caption{(a) Interactions between aggregate atoms and environment atoms, where $V_{rs}$ and $V_{rp}$ represent van der Waals interactions of environment atoms with aggregate atoms in $s$ and $p$ states, respectively and $W$ represents dipole-dipole interactions between aggregate atoms. (b) EIT-scheme for environment atom, where $\Omega_p$ and $\Omega_c$ are probe and coupling Rabi frequencies, respectively. $\Delta_p$ and $\Delta_c$ are detunings while $\Gamma_p$ is the decay rate from $\ket{e}$ on the probe transition. (c) Microwave coupling of the Rydberg $s$ state to the $p$ state in the aggregate, with Rabi frequency $\Omega_{mw}$ and detuning $\Delta_{mw}$.
\label{EIT_level_scheme}}
\end{figure}

For this we consider a few additional environment atoms, separately trapped at specific locations $\mathbf{x}_n$ in the vicinity of the aggregate atoms.
The environment atoms are then optically coupled from their ground state to two other states in the scheme of electromagnetically induced transparency (EIT) involving an auxiliary Rydberg state $\ket{r}$, shown in \frefp{EIT_level_scheme}{b}. By populating the $\ket{r}$ state the $\ket{r}\rightarrow\ket{s,p}$ interactions introduce on-site static energy disorder, 
which implies that the energy of a certain system state then depends on \emph{which} atoms are in the $\ket{s}$ or $\ket{p}$ state, rather than just their total populations. Additionally, when environment atoms reach the decaying state $\ket{e}$, the resultant light absorption allows the discrimination of aggregate states $\ket{s}$, $\ket{p}$ and hence causes measurement induced dephasing. For a detailed discussion of these effects, we refer to \cite{schoenleber:immag}. 
The complete many-body Hamiltonian for the assembly of Rydberg aggregate plus environment atoms is described in \eref{Htot_app}-\eref{vander_interaction}. Calculating the corresponding time evolution of this open quantum system while explicitly including the non-trivial environment degrees of freedom is a formidable task. However, for the selected parameter regimes that we focus on below, it is possible to simplify the numerical treatment considerably.
These are the cases in which it is possible to adiabatically eliminate the dynamics of environment atoms using techniques of \cite{sorensen:adiabelim} as in \cite{schempp:spintransport,schoenleber:immag}, which requires relaxation that is fast compared to the timescale for dipole-dipole interactions, $\Gamma_p \gg $max$_{nm}[W_{nm}]$.
In that regime one can obtain the effective master equation
 \begin{eqnarray}
\dot{\hat{\rho}}  = -i[\sub{\hat{H}}{agg} +\sub{\hat{H}}{mw}  + \sub{\hat{H}}{eff} ,\hat{\rho}] +\sum_\alpha {\cal L}_{\sub{\hat{L}}{eff}^{(\alpha)}}[\hat{\rho}].
\label{mastereqn}
\end{eqnarray}
where the super-operator  ${\cal L}_{\hat{O}}[\hat{\rho}]=\hat{O}\hat{\rho}\hat{O}^{\dagger}-(\hat{O}^{\dagger}\hat{O}\hat{\rho}+\hat{\rho}\hat{O}^{\dagger}\hat{O})/2$ accounts for the decoherence in the system. The density matrix is written as $\hat{\rho} = \sum_{kl} \rho_{kl} \ket{\chi_k}\bra{\chi_l}$, where $\ket{\xi_k}$ is a many body state for the Rydberg aggregate only, given in the form of a tensor product
\begin{eqnarray}
\ket{\chi_k}= \bigotimes_{i=1}^N \ket{T_i^k}=\ket{T_1^k T_2^k ...T_N^k},
\label{manybodystates}
\end{eqnarray}
where $T_i^k \in \{s,p\}$ and $k \in \{1,2,...2^N\}$ as we have $2^N$ states for $N$ aggregate atoms.

In \eref{mastereqn}, $\sub{\hat{H}}{agg}$, $\sub{\hat{H}}{mw}$, $\sub{\hat{H}}{eff}$ and $\sub{\hat{L}}{eff}^{(\alpha)}$ are the aggregate Hamiltonian \eref{Hagg}, microwave Hamiltonian \eref{Hmw}, effective Hamiltonian and Lindblad jump operator, respectively. The index $\alpha$ numbers the environment atoms used to control decoherence.
We can write 
\begin{eqnarray}
\sub{\hat{H}}{eff}&=&\sum_k  \sub{h}{eff}^{(k)} \ket{\chi_k}\bra{\chi_k},
\label{main_Heff}
\\
\sub{\hat{L}}{eff}^{(\alpha)}&=&\sum_k  \sub{\gamma}{eff}^{(k,\alpha)} \ket{\chi_k}\bra{\chi_k},
\label{main_Leff}
\end{eqnarray} 
with details given in \aref{app_LeffHeff}. 

Both, $\sub{\hat{H}}{eff}$ and $\sub{\hat{L}}{eff}^{(\alpha)}$ depend on the probe and coupling Rabi frequencies in the EIT scheme implemented for the environment atoms, $\Omega_p$ and $\Omega_c$ respectively, the corresponding detunings $\Delta_p$ and $\Delta_c$, as well as the spontaneous decay rate of the intermediate state $\Gamma_p$. Crucially, they also depend on the van-der-Waals interactions between environmental atoms and aggregate atoms. All of these can be varied in experiments, rendering both disorder and dephasing distributions widely tunable.

Later we shall engineer disorder and dephasing in this setup through a synthetically controlled distribution $p_\mathbf{x}$ of locations of environmental atoms. 
These can for example be chosen from a Gaussian random distribution, with a fresh disorder realization in each experiment. This variation in locations of the environment atoms leads to variations of the aggregate matrix elements $\sub{h}{eff}^{(k)}$ and matrix elements in the effective Lindblad operator $\sub{\gamma}{eff}^{(k,\alpha)}$, as we shall see in \sref{inhomogen}.

\section{Two-dimensional Rydberg spectra}
\label{results}

Here we show that with Rydberg atoms and microwave pulses one can simulate the basic features of 2D optical spectroscopy of molecular aggregates.
For the example of small aggregates and a specific choice of environmental parameters, we will discuss relevant aspects of the obtained spectra such as peak broadening caused by dephasing and static disorder. 
We will then discuss aspects specific to an implementation in Rydberg experiments.

\subsection{Exemplary parameters}
\label{parameters}

For the following demonstrations, we use the parameters summarized in \tref{parameter_table}. 
The different parameter sets are chosen on one hand to demonstrate various aspects of 2D spectroscopy and on the other hand to allow tractable numerical and analytical analysis: the numerically simpler equation \bref{mastereqn} is valid, and even diagonal when expressed in the basis of the eigenstates of the effective Hamiltonian,  as discussed in \aref{photon_echo}.
The former aids numerical solutions, since the more general equation \bref{generic_evolution_equation} would be very time-consuming to solve, and latter allows simple analytical results.
However, importantly, 2D spectroscopy of embedded Rydberg aggregates can also be performed for a broad range of parameters well outside the range of validity of \bref{mastereqn} that are known to lead to more interesting dynamics \cite{genkin:markovswitch,schoenleber:thermal}, which is however challenging for numerical or analytical considerations. 

For the parameters in the table, the order of magnitude of the most energetic eigenstate will be $|\sub{U}{max}|\sim $max$_{nm}[W_{nm}]$, with $\sub{U}{max}\approx5$ MHz. According to the discussion in \sref{microwave} we thus need pulse-lengths $\delta t \lesssim 200$ns, which are technically feasible. Additionally, since we do not include the decay of the Rydberg excitation, the longest possible pulse sequence of duration $\sub{\tau}{tot}$, must be within the Rydberg system lifetime $\sub{\tau}{eff}\approx\tau_0/N=14.6$ $\mu$s , where $\tau_0=44$ $\mu$s is the lifetime of a single atom in $s$-state\footnote{Lifetimes in $p$-states are larger, so this is a worst case estimate.} and $N=3$.

\subsection{Rydberg dimer spectra}
\label{dimerspec}

\begin{figure}[t]
\centering
\includegraphics[width=0.99\columnwidth]{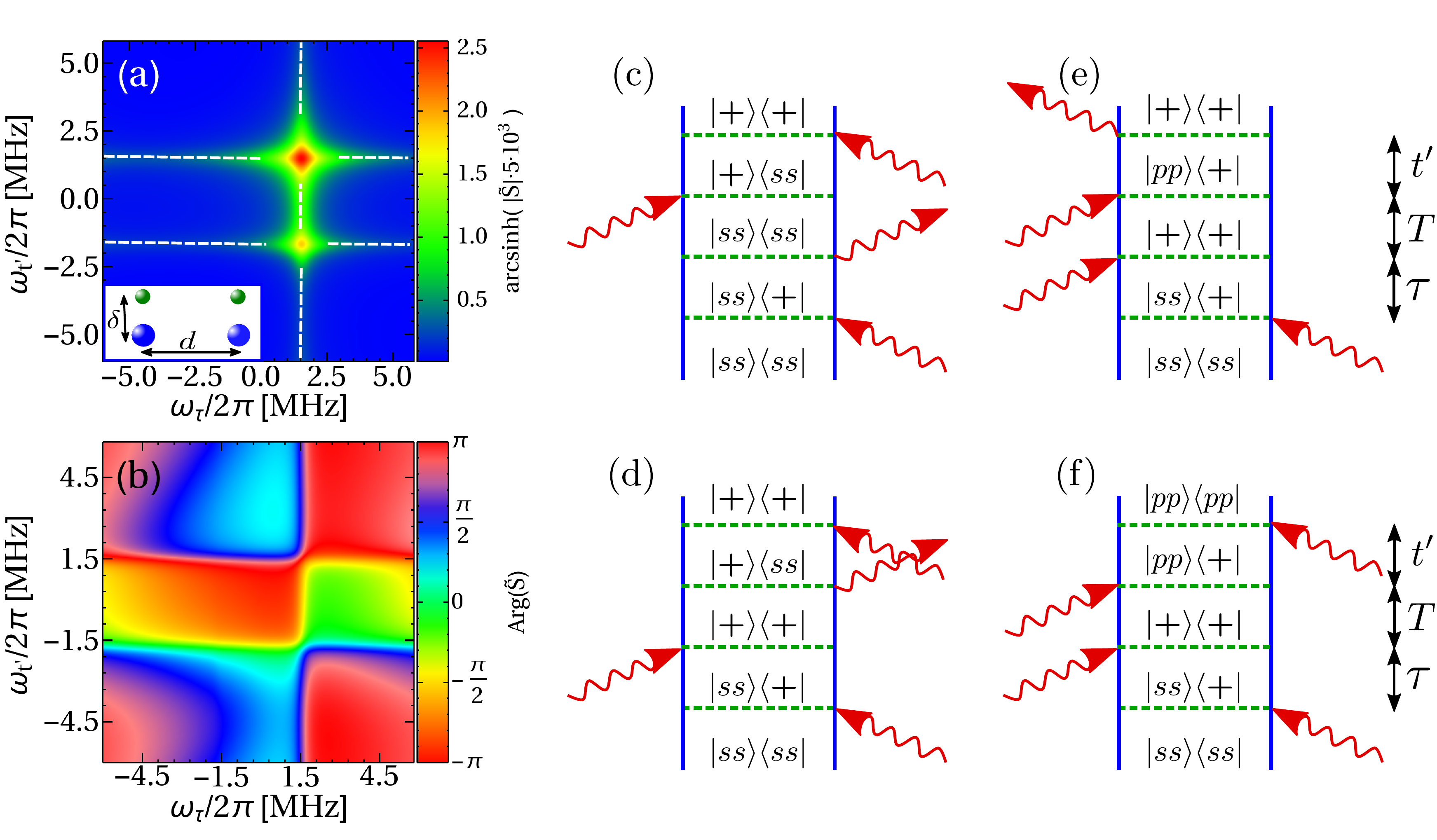}
\caption{Two dimensional spectrum of a monitored Rydberg dimer for waiting time $T=0$. The atom arrangement is shown in the inset of (a) with $d=10$ $\mu$m (resulting in an interaction $W(d)/(2\pi)=1.619$ MHz, $\delta=1.5$ $\mu$m, with all other parameters shown in \Tref{parameter_table}. Blue balls indicate aggregate atoms, and green ones environment atoms.  The zero of all frequency axes corresponds to the central microwave frequency $\omega_0$. (a) Absolute value of the signal intensity $\tilde{S}(\omega_\tau,\omega_{t'})$. White dashed lines are 1D cuts for which we compare numerical and analytical results in \fref{comparison_pic}. (b) the phase arg$[\tilde{S}]$. (c)-(e)
Double sided Feynman diagrams \cite{mukamel2000m} representing the dominant Liouville space pathways for the spectra shown in (a) and (b). Curly incoming (outgoing) arrows representing interaction of the Rydberg dimer with a microwave pulse resulting in (de-) excitation.
\label{2dspectra_dimer}
}
\end{figure}

We first consider the simplest example: Two Rydberg atoms separated by a distance $d=10$ $\mu$m form the aggregate, shown as blue balls in the inset of \fref{2dspectra_dimer}{a}.  The dipole-dipole interaction Hamiltonian  \bref{Hagg}  is then given by 
\begin{eqnarray}
\sub{\hat{H}}{agg}=\left[
\begin{array}{cccc}
2 \epsilon_s  &0 &0 &0\\
0  &\epsilon_s+\epsilon_p &W(d) &0\\
0  &W(d) &\epsilon_s+\epsilon_p &0\\
0  &0 &0 &2\epsilon_p
\end{array} \right],
\label{dimer_hamil_matrix}
\end{eqnarray}
where the matrix representation is consecutively in terms of the two-body basis states $\{\ket{ss},\:\ket{sp},\:\ket{ps},\:\ket{pp} \}$. Diagonalisation of \bref{dimer_hamil_matrix} yields a non-excited state $\ket{ss}$, with energy $2 \epsilon_s$, two single-exciton states $\ket{\pm}=(\ket{ps}\pm\ket{sp})/\sqrt{2}$ (with energies $2 \epsilon_s+ \hbar\omega_{sp}\pm W(d)$) and one two-exciton state $\ket{pp}$ (with energy $2 \epsilon_s+2\hbar \omega_{sp}$). Here $\hbar \omega_{sp}=\epsilon_p-\epsilon_s$ as defined earlier.
 
Each aggregate atom is flanked by a detector atom placed $\delta =1.5$ $\mu$m away, sketched as green balls in \fref{2dspectra_dimer}{a}.  This introduces additional energy shifts of the above states, as discussed in \aref{app_perttheory}, which are  however minor corrections for the present parameters.

\begin{center}
	\begin{table}
		\begin{tabular}{| c| c | c| c| c| c | c| c|} 
			\hline
			Figure & $C_{4,rp}/(2\pi)$ & $C_{6,rp}/(2\pi)$ & $C_{6,rs}/(2\pi)$ & $\Omega_p/2\pi$  & $\Omega_c/2\pi$  & $ \Delta_p/(2\pi)$ & $\Delta_c/(2\pi)$ \\ 
			&[MHz $\mu$m$^4$] &[MHz $\mu$m$^6$] & [MHz $\mu$m$^6$] & [MHz] & [MHz] & [MHz] & [MHz]\\
			\hline
			\bref{2dspectra_dimer}  & $-1032 $ & & $-87$   & $1.8$  & $30$ &  $0$ & $0$ \\
			\bref{2dspectra_trimer}(a)  & $-1032$  & & $-87$ & $0.3$  & $10$ &  $0$ & $0$\\
			\bref{2dspectra_trimer}(b)  & $-1032$  & & $-87$ & $1.3$  & $30$ &  $0$ & $0$ \\
			\bref{disorder}(a)  & & $-0.4$  & $-0.1$ & $40$  & $120$  &  -$20$ & $20$ \\
			\bref{disorder}(b)  & & $-0.4$  & $-0.1$ & $30$  & $120$  &  -$40$ & $40$ \\
			\hline 
		\end{tabular}
			\caption{ \label{parameter_table} Parameters used in calculations of 2D spectra in the following figures. All simulations use  $C_3/(2\pi)=1619$ [MHz $\mu$m$^3$], $\Gamma_p/(2\pi)=6.1$ MHz and $\Omega_{mw}/(2\pi) = 4$ MHz.}
	\end{table}
\end{center}
Using phase cycling as described in \sref{signal_and_cycling} with details listed in \aref{app_phasecycling}, we show the absolute value of the expected 2D spectrum and its complex phase in \fref{2dspectra_dimer}.  
Here and in the following we plot the absolute value using  ${\rm arcsinh}(5\cdot10^3\, |\tilde{S}|) $  to make small features more visible.
We have obtained the spectrum shown in \fref{2dspectra_dimer} by numerically simulating \eref{mastereqn} for 64 different time delays in $\tau$ and $t'$ each.
Simulations employ the high-level package XMDS \cite{xmds:paper,xmds:docu}. For the pulses we used smoothened step functions with rise-time $\sub{t}{rise}\ll \delta t$ \footnote{The specific pulse shape we use is $A(t_0)=\frac{1}{2}(\tanh[(\sub{t}{eff}-|t_0|)/\sub{t}{rise}]+1)$, if ($-\frac{\delta t}{2} < t_0 \leq \frac{\delta t}{2}$), with $\sub{t}{eff} = \frac{\delta t}{2} - 3\sub{t}{rise}$, $\sub{t}{rise}=0.01 \mu$s and $\delta t=0.05 \mu$s.}.

We also show in \ref{app_phasecycling} how the spectrum  looks for a specific single pulse sequence.
In these calculations, we have made sure to stay in regime of very low $p$-excitation probability and thus low signal strength, such that perturbation theory remains valid. In practice, one would want to venture to the limit of the perturbative regime with higher probability of $p$-excitation, in order to avoid the need for too many repetitions of the experiment, see \sref{sec_ensemble}.

For this simple case of a symmetric dimer, we can easily understand the position, intensity and lineshape of peaks in the spectrum from perturbation theory \cite{mukamel:book}.
One can identify and represent the relevant contributions to the signal by so-called double-sided Feynman diagrams, shown in \fref{2dspectra_dimer} and discussed in \aref{app_perttheory} and \aref{photon_echo}. Of crucial importance are the energy differences and transition dipoles between eigenstates of $\sub{H}{agg}+\sub{H}{eff}$, see \eref{mastereqn}, which are listed in \tref{dimer_evals}. Using the notation of \frefp{sketch}{c}, we find that from the initial ground-state $\ket{ss}$ transitions only proceed via the states $\ket{ss}\leftrightarrow \ket{+}\leftrightarrow \ket{pp}$.  The corresponding frequencies are $\omega_{+,ss}= (2\pi)\times1.56$ MHz and $\omega_{+,pp}=  (2\pi)\times 1.67$ MHz,  relative to $\omega_0$ and transition strengths $\mu_{+,ss}=\mu_{+,pp}=\sqrt{2} \mu$. The state $\ket{-}$ is not involved due to its anti-symmetry. The first interaction creates a coherence between the ground state $\ket{ss}$ and $\ket{+}$.
For the parameters used, the environment does not induce coupling between different elements of the reduced density matrix of the system.
Therefore, during the time interval $\tau$, the system accumulates a phase $\omega_{+,ss} \tau$. 
The Fourier transformation with respect to $\tau$ gives a peak at the position  $\omega_\tau=\omega_{+,ss}$.
The width of this peak is determined by the corresponding dephasing rate $\Gamma_{+,ss}$.
In order to provide a contribution to the signal, which is the number of p-excitations, the second pulse must bring the system either to the population of the $\ket{+}$ state (diagrams c-e) contributing a single p, or to the doubly p-excited population (diagram f).
 For the present choice of parameters the environment does not affect populations.
 For diagram (c) and (d), the system is again in the coherence between ground state $\ket{ss}$ and $\ket{+}$ during the time-evolution $t'$, after the third pulse. As before, the Fourier transform then gives a peak at $\omega_{t'}=\omega_{+,ss}$.
 Thus diagrams (c) and (d) give rise to the diagonal peak at $ (2\pi)\times(1.56 , 1.56)$ MHz.
 Its intensity is proportional to $|\mu_{+,ss}|^4=4 \mu^4$. 
In contrast, for diagrams (e) and (f), the third pulse brings the system into a coherence between the doubly excited state $\ket{pp}$ and $\ket{+}$, during the time-evolution $t'$.
Since the frequency $\omega_{pp,+}$ is different from the frequency $\omega_{+,ss}$, see \frefp{sketch}{c}, this results (after Fourier transformation) in an off-diagonal peak at $(2\pi)\times(1.56, -1.67)$ MHz.
The complete spectrum for the symmetric dimer can to a good approximation be written analytically as 
\begin{eqnarray}
\label{dimer-sig}
S(\omega_\tau,\omega_t')&=-|\mu_{+,ss}|^4\frac{E_0^4}{(\omega_\tau-\Omega_{+,ss})(\omega_{t'}-\Omega_{+,ss})}\nonumber\\
&+|\mu_{+,ss}|^2|\mu_{pp,+}|^2\frac{E_0^4}{(\omega_\tau-\Omega_{+,ss})(\omega_{t'}-\Omega_{pp,+})}
\end{eqnarray}
with $\Omega_{a,b}=\omega_{a,b}-i \Gamma_{a,b}$.  For details see \aref{app_perttheory}-\aref{feynman_diag}.
Here, in particular, infinitely short pulses have been assumed. For the present choice of parameters and a single environment atom near each site, we find $\Gamma_{+,ss}\approx(2\pi)\times0.2155$ MHz and $\Gamma_{pp,+}\approx(2\pi)\times0.2150$ MHz. 
There is very good agreement between the full non-perturbative calculation and the analytic result, as can be seen in \fref{comparison_pic}.
By changing the parameters of lasers acting on the environment atoms, our system allows tuning the widths $\Gamma_{+,ss}$ and $\Gamma_{pp,+}$.
Note, that in parameter regimes where the simple master equation \bref{mastereqn} is no longer valid the environment can cause more complicated effects than simple dephasing.

If one introduces asymmetry to the system, for example by using a different placement of environment atoms or a different number of them near each of the two aggregate atoms, the $\ket{-}$ state is no longer perfectly antisymmetric and additional weak contributions from this transition arise,  which would give rise to additional peaks in the spectrum.

\subsection{Rydberg trimer spectra}
\label{trimerspec}
\begin{figure}[t]
	\centering
	\includegraphics[width=0.99\columnwidth]{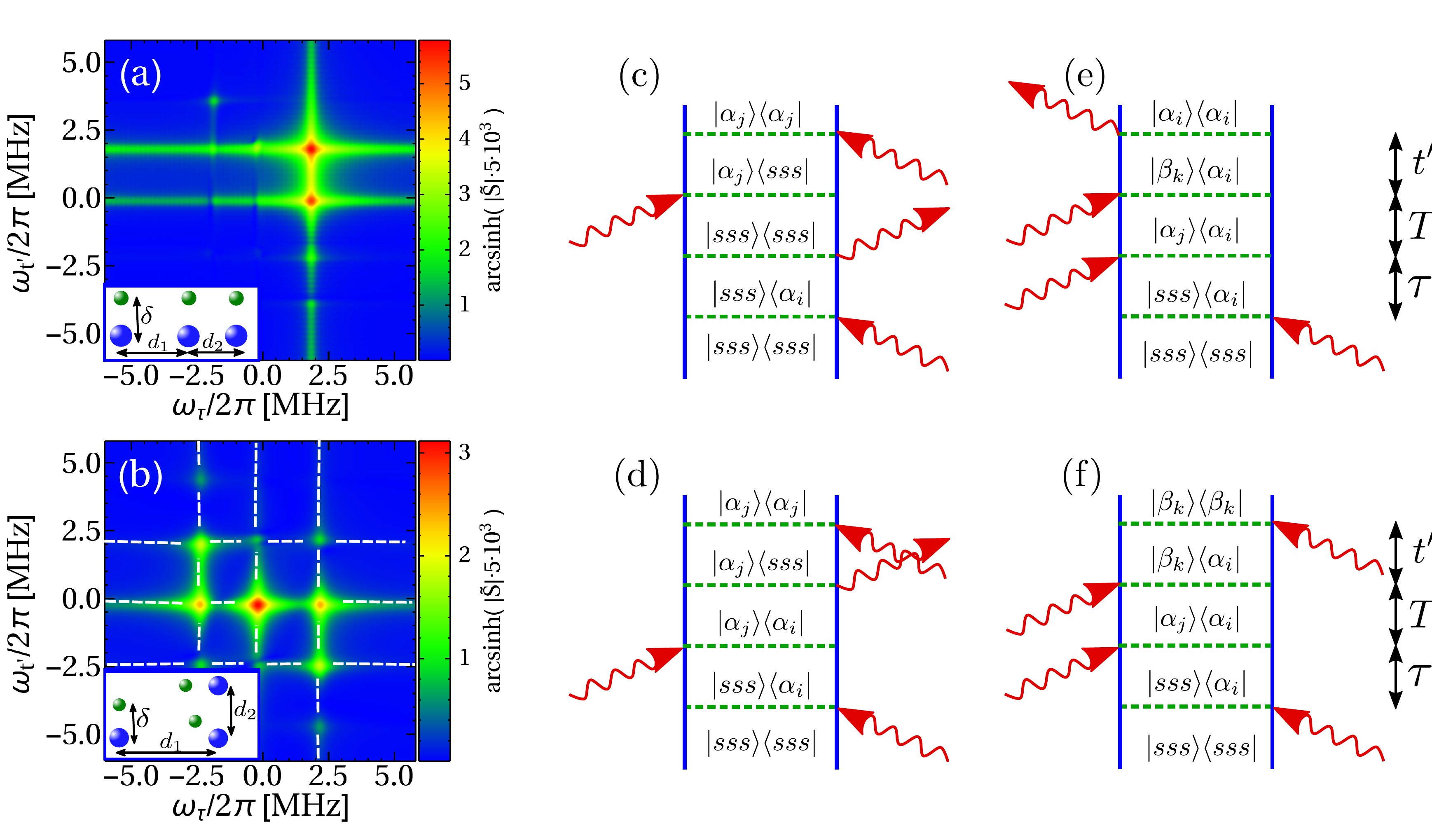}
	\caption{Two dimensional spectrum of an environmentally monitored Rydberg trimer with geometry shown in the insets. Blue balls indicate aggregate atoms, and green ones environment atoms. (a) non-equidistant trimer with $d_1=10$ $\mu$m and $d_2=12.6$ $\mu$m, $\delta=2.5$ $\mu$m and (b) L-shaped trimer with $d_1=10$ $\mu$m and $d_2=12.6$ $\mu$m, $\delta=1.5$ $\mu$m and all other parameters are shown \Tref{parameter_table}. (c)-(e) Doublesided Feynman diagrams representing the relevant Liouville space pathways for the rephasing signal for the Rydberg trimer.
		\label{2dspectra_trimer}
	}
\end{figure}

%
The Rydberg trimer is a significant extension compared to the dimer, as it now allows different geometries ranging from ones that lead to only one  bright state (as in the previous dimer case) to cases where all transitions are dipole-allowed.
In the following we show an example for each, but focus on the latter. As in the previous section for the dimer, we first also take one environment atom per site. 
These atoms are arranged as shown in the insets of \fref{2dspectra_trimer}. 

In the first case, of panel (a), the trimer is arranged linearly, but slightly non-equidistant. 
This gives rise to several dipole-allowed transitions, which can be seen by the emergence of new peaks.
These peaks are however very small compared to those involving the dominant symmetric state, which behaves akin to the dimer case in \sref{dimerspec}.

To obtain a richer spectrum we focus on the L-shaped arrangement shown in the inset of \fref{2dspectra_trimer}{b}.
Here, because of the anisotropy of the dipole-dipole interactions, the eigenstates have a very different form compared to the linear arrangement, which results in the emergence of several transitions that have comparably large strength.
The spectrum in \frefp{2dspectra_trimer}{b} thus shows a multitude of diagonal and off-diagonal peaks.
As for the dimer case their positions, intensities and lineshapes can be understood from perturbation theory.
The peak positions are again determined by the eigenenergy differences of the effective Hamiltonian in \bref{effhamil_SE_appendixC}.  Besides the distances of aggregate atoms, these are also strongly affected by the relative orientation of aggregate atoms in the 2D plane, since we consider anisotropic dipole-dipole interactions here. Finally there are some minor additional offsets, stemming from interactions with the environment atoms.
We will comment here only on the main features.
The trimer has three single-exciton states and three double-exciton states. This gives rise to three transitions between ground state $\ket{sss}$ and the single excited states and 9 transitions between single-exciton states and double exciton states. The corresponding frequencies are given in \tref{transition_table_trimer}.
Similar to the discussion in the previous section, the relevant contributions arising during the first time interval $\tau$ stem from a coherence between ground and single exciton states.
After Fourier transform these show three respective transition frequencies, indicated by the three vertical lines in \frefp{2dspectra_trimer}{b}. 
The same frequencies can also appear during the time-evolution $t'$, indicated by lines when also $\omega_{t'}$ equals one of these. 
The discussion so far results in the three diagonal peaks at $(2\pi)\times(-2.29, -2.29)$, $(2\pi)\times(-0.09,-0.09)$, $(2\pi)\times(2.29,  2.29)$ MHz.
During the second time interval $t'$, the system can instead also be in coherences between a singly excited state and a doubly excited state. This gives three additional contributions which result in the eight off-diagonal peaks in \frefp{2dspectra_trimer}{b} located at $(2\pi)\times(-2.29,-0.03)$, $(2\pi)\times(-2.29,4.55)$, $(2\pi)\times(-2.29,2.17)$, $(2\pi)\times(-0.09,-2.23)$, $(2\pi)\times(-0.09,2.35)$, $(2\pi)\times(2.29,-4.61)$, $(2\pi)\times(2.29,-0.03)$ and $(2\pi)\times(2.29,-2.41)$ MHz. A ninth peak at $(2\pi)\times(-0.09,-0.03)$ MHz merges with the central diagonal peak.
See also \tref{transition_table_trimer} in~\aref{trimer}, where we list all relevant system parameters and eigenstates.
%
\subsection{Separating homogeneous and inhomogenous broadening}
\label{inhomogen}
%
\begin{figure}[t]
	\centering
	\includegraphics[width=0.9\columnwidth]{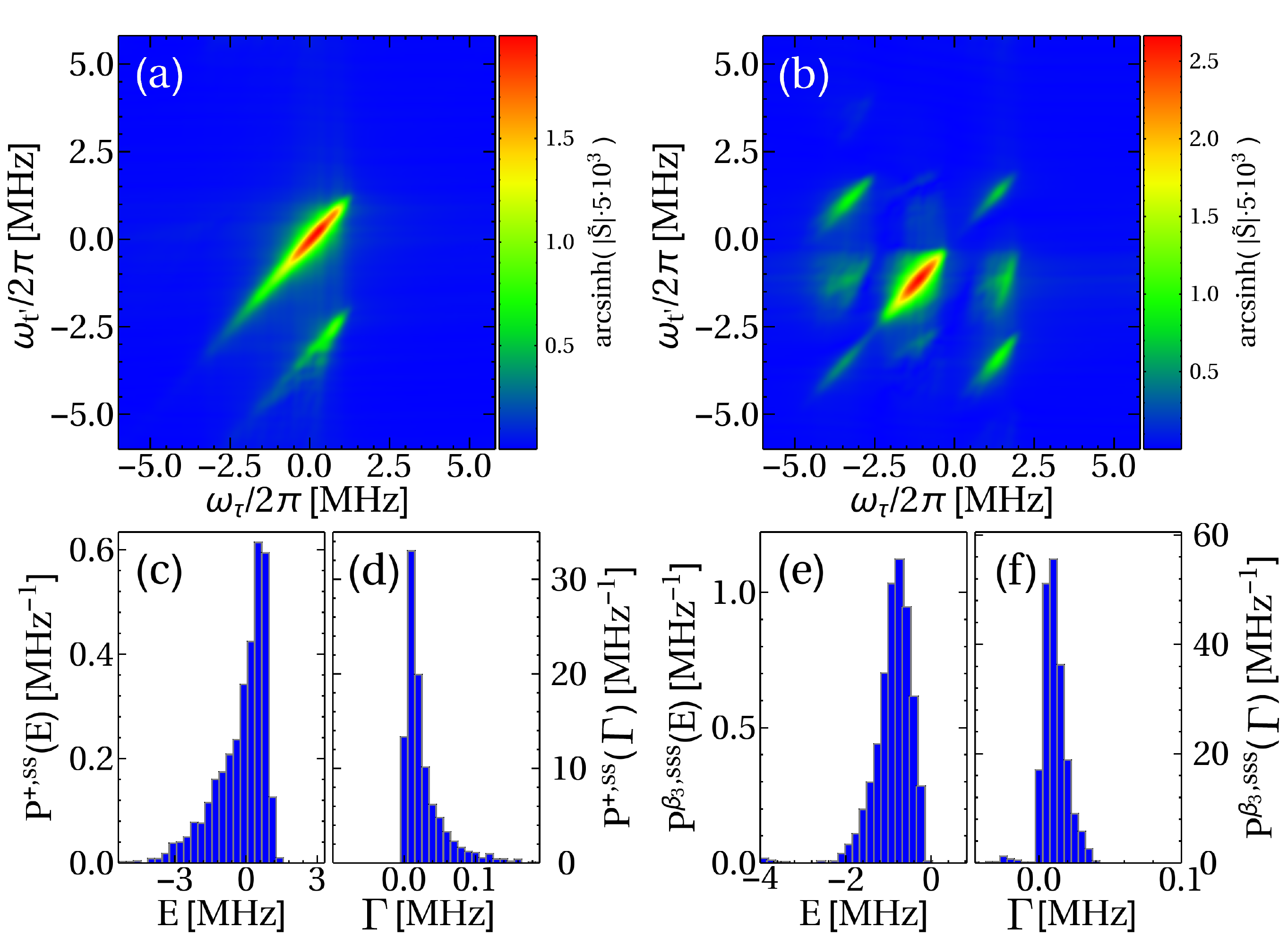}
	\caption{Two dimensional spectrum of an environmentally monitored Rydberg aggregate, with larger energy disorder compared to dephasing, atom color coding as in earlier figures. (a) Rydberg dimer;  $d=10$ $\mu$m and two environment atoms per aggregate atom. (b) Rydberg trimer;  $d_1=10$ $\mu$m, $d_2=12.6$ $\mu$m, see inset \frefp{2dspectra_trimer}{b}.
All parameters are given in \Tref{parameter_table}. Each aggregate atom is flanked by two environment atoms, the trapping location of which is synthetically disordered independently, with width $\sigma = 565$ nm along the indicated directions, see text. We average $1000$ realizations in each case. (c) Resultant histogram for the distribution of transition energies  $P(E)$ and (d) dephasing rates $P(\Gamma)$ (defined in the text) on the relevant $\ket{ss}\leftrightarrow\ket{+}$ transition, for the same parameters as in (a). (e,f), the same for parameters of panel (b),  focussing on the $\ket{sss}\leftrightarrow\ket{\beta_3}$ transition, see \ref{trimer}.
\label{disorder} 
	}
\end{figure}

A strength of higher dimensional spectroscopy is the disentangling of different types of line broadening mechanisms. This can be clearly illustrated and studied in the embedded Rydberg setup, where one can independently tune the two types of broadening \cite{schoenleber:immag,schempp:spintransport,genkin:markovswitch,schoenleber:thermal}.

As stated earlier, we assume all atoms to be tightly trapped in the harmonic oscillator ground-state of their respective trap. Detector atoms should be easily trappable against the weak, dressed van-der-Waals Forces exerted on them, while aggregate atoms might have to undergo additional ground-state cooling after each phase set, or sequence, since they experience much stronger bare dipole-dipole interactions. If we assume ground-state trapping, one can envisage an effective Hamiltonian, obtained after integrating out spatial degrees of freedom.
This would not contain any disorder without experimental intervention.  However disorder can be introduced and controlled synthetically, by moving selected optical trap centres after measuring one complete spectrum, which we assume to happen in the following. 

We  now explore averaged spectra for a trimer as in the previous section, but with parameters adjusted to enhance diagonal disorder, arising via $\sub{\hat{H}}{eff}$ in \eref{main_Heff} compared to dephasing, arising from \bref{main_Leff}. Synthetical disorder is included by randomly varying the trapping location of two environment atoms per aggregate atom, after each spectrum acquisition. By ``spectrum acquisition" we imply gathering the complete dataset: measuring the mean signal strength (p-population) for all sets of phases and all required time-delays. The detector traps in realisation number $k$ are then at $\mathbf{x}_k=\mathbf{x}_0 + \sigma \eta_k\mathbf{n}$, where the central positions $\mathbf{x}_0$ are described in the caption of \fref{disorder}, $\mathbf{n}$ are normal unit vectors along the axes indicated by black arrows in the insets, $\eta_k$ is drawn from a Gaussian distribution of unit variance, and $\sigma$ thus sets the width (standard deviation) of this position distribution.

 As mentioned in \sref{environment}, the variation in locations of the environment atoms leads to variations of matrix elements in the effective Hamiltonian $\sub{h}{eff}^{(k)}$ and effective Lindblad operator $\sub{\gamma}{eff}^{(k,\alpha)}$ see \ref{app_LeffHeff} for details. These in turn affect the energies and dephasing rates of transitions between system eigenstates $\ket{k}$ and $\ket{k'}$. We define $P^{k'k}(E)$ as the probability for the resultant shifted $k\leftrightarrow k'$ transition energy to become $E$, and correspondingly $P^{kk'}(\Gamma)$ as the probability for the dephasing rate of that transition to be $\Gamma$. The latter is given by the real part of \eref{eq:Gammkk} in \ref{app_LeffHeff}.

 In \frefp{disorder}{a,b} we show examples of spectra with variations of the environment-atom positions on the order of $500\,{\rm nm}$. 
The corresponding distributions of the relevant transition energies and dephasing rates are provided in panels {c}-{f}.
More details can be found in the supplemental information, where we also provide distributions for the individual matrix-elements of $\sub{\hat{H}}{eff}$.
 The shift, width and shape of 2D peaks along the diagonal reflects the disordered distribution of transition energies in \frefp{disorder}{c,e}, defined as $P^{k'k}(E)$ above. In contrast, the width in the off-diagonal direction matches the dephasing distributions $P^{k'k}(\Gamma)$ in \frefp{disorder}{d,f}. We further explicitly verified this allocation by modelling the system in \fref{disorder} with tenfold increased dephasing terms, i.e. real part of Lindblad operator in \eref{mastereqn}, and found that indeed only the anti-diagonal width of the peaks reflects this.
 
\subsection{Spectrum acquisition time and ensemble spectroscopy}
\label{sec_ensemble}

\begin{figure}[t]
	\centering
	\includegraphics[width=0.9\columnwidth]{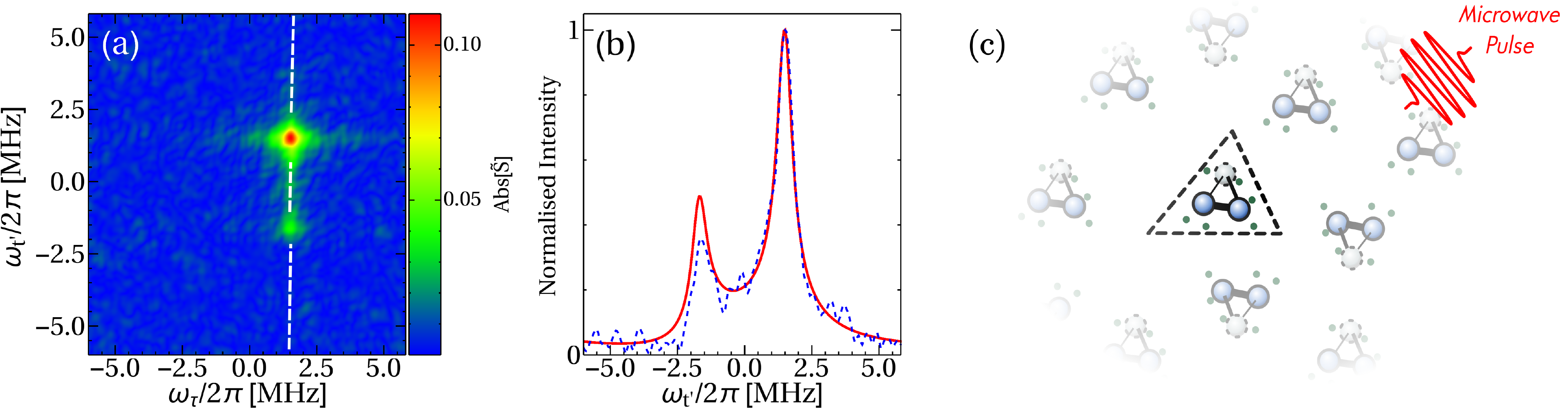}
	\caption{Simulation of 2D-spectrum in the presence of discrete atoms counts. (a) Two dimensional spectrum of embedded Rydberg dimer, as in \fref{2dspectra_dimer}, but obtained by explicitly simulating the detection of a discrete number of atoms, assuming the measurement is repeated $\sub{N}{rep}=10^5$ times and $\Omega_{mw}/(2\pi) = 16$ MHz. (b) Comparison along the value of $\omega_\tau$ indicated by the white dashed line in (a) with the analytical expectation. (c) To accelerate obtaining the average final Rydberg $p$ population, we propose the interrogation of a large number of replica aggregates in parallel by the same microwave pulse.
	\label{fig:Nrep}
	}
\end{figure}

To experimentally measure any of the spectra shown, one has to gather enough statistics to obtain a sufficiently high signal to noise ratio. Since the signal is of fourth order in the weak electric field, the probability to promote atoms to the p-state is quite small. Now, phase cycling relies on a precise cancellation of an undesired large leading order signal in the sum \bref{cycledsignal}. Hence, we determine the signal to noise ratio required for this cancellation to work.
We explicitly explored this aspect in the simulations shown in \fref{fig:Nrep} and discuss it in more detail in \ref{ensemble_calc}. 
For the parameters of \fref{2dspectra_dimer}, we infer the need for about  $\sub{N}{rep}\approx10^5$ repetitions of each measurement, where the latter refers to a single sequence of pulses with subsequent measurement of the total p-population.

Based on this one can estimate the duration of acquisition of a single spectrum.  The maximum interval between pulses, $\sub{\tau}{max}$ is linked via $\sub{\tau}{max}=\pi/\Delta \omega$ with the desired frequency resolution $\Delta \omega$. In order to resolve the shape of peaks, $\Delta \omega$ ought be smaller than typical dephasing rates $\gamma$ in the system, which govern the width of peaks. For the example in \sref{dimerspec} this yields $2\sub{\tau}{max}\approx10$ $\mu$s.
Let the duration of the two pulses be $\tau= n \Delta t$ and $t'= m \Delta t$, where the minimal time-step size $\Delta t=\pi/\sub{\omega}{max}$, as follows from discrete Fourier theory. 
In turn the maximum frequency $\sub{\omega}{max}$ of the target spectrum has to exceed all relevant eigenenergies of the system.
The total duration for one sequence of pulses will then be 
$\sub{\tau}{tot} = 4 \sum_{n,m}^{\sub{N}{pts}/2} (n+m)\Delta t$, where the number of points per frequency axis, $\sub{N}{pts}$, is determined by $\Delta \tau$ and $\sub{\tau}{max}$ (and similar for the $t'$ axis). The expression gives  $\sub{\tau}{tot}= \Delta t \sub{N}{pts}^2 (1 + \sub{N}{pts})$, and when including the number of repetitions discussed above and the number of phases  $\sub{N}{phases}$ to be cycled over, the total duration is approximately
\begin{eqnarray}
 \label{expt_duration}
\sub{T}{tot}&=\sub{T}{rep} \sub{N}{rep}\equiv \frac{\pi \sub{N}{pts}^2(\sub{N}{pts}+1)}{\sub{\omega}{max}}\times\sub{N}{phases}\times \sub{N}{rep}.
\end{eqnarray}
We note that this is a lower limit, since we do not consider any time required for the initialisation of each measurement, such as trap loading.
Assuming that a single repetition takes $\sub{T}{rep} \approx 0.58$ seconds, for the exemplary parameters of Fig.~\ref{fig:Nrep} we find $\sub{T}{tot} \approx$ 16 hours.
 Recall, however, that our parameters are largely chosen to ease theory, and we expect that the total time can be lowered by a significant factor when using different parameters such as higher principal Rydberg quantum numbers up to~$n\approx 80$. Higher principal quantum numbers increase the dipole-dipole interaction strength $C_3$ in \bref{dipole_dipole_int} according to $C_3\sim n^4$, and consequently the system eigenenergy differences ${\cal E}_n-{\cal E}_m$. The 2D spectrum of interest then covers a larger frequency range, allowing less frequency resolution and hence much shorter delay-times, if the relative widths of peaks are preserved.  
 
 Other possibilities of reducing the experimental time are techniques based on reducing the overall number of required samples, such as
are 1-norm optimisation \cite{Katz_compressedGhosts_APL,Almeida_2Dcompsamp_JPCL,Andrade_compressedsensing_PNAS} or sparse sampling optimized with genetic algorithms \cite{Roeding_sparsesamp_JCP}.

For setups where the above duration is nonetheless a hindrance, we propose the simultaneous interrogation of an ensemble of identically prepared replica of the Rydberg aggregate  as sketched in \frefp{fig:Nrep}{c} by the same microwave pulse. Replicas should be far enough separated not to affect one another. Such a system can be realised using spatially structured laser fields for trapping the atoms in arbitrary 2D geometries \cite{nogrette:hologarrays,wang_tweezerarray}.
As discussed in \sref{ryd_aggregate}, our signal is simply the total number of Rydberg atoms in the $p$-state, hence simultaneous interrogation would automatically average over all aggregates in the system. 
For the discussion above, one can thus significantly reduce the number of repetitions $\sub{N}{rep}$ for the ensemble average, since many repetitions would be provided already by a single experimental run.  A challenge of the replica setup would be the simultaneous laser addressing of a large number of environmental atoms, which can be tackled by working in a 2D plane, and widening EIT laser foci into laser sheets. Residual inhomogeneities of Rabi frequencies would then simply contribute to the disorder distribution.

In these simulations and those of the preceding section, both detector and aggregate atoms are expected to be at a precisely determined trapping location. We explore in \aref{app_disorder_versus_cycling} what happens if these locations themselves vary \emph{during} acquisition of a single spectrum, and find that trapping centre fluctuations of spatial widths exceeding $\sigma = 0.2$ $\mu$m begin to require an increasing number of repetitions.

\section{Conclusions and outlook}
\label{conclusions}

We have outlined a protocol for non-linear two-dimensional spectroscopy, which is a powerful tool in material science and physical chemistry, applied to the realm of ultracold Rydberg physics.
Through the excellent control over the latter, this would enable a benchmark platform for the further development of algorithms and tweaks for multidimensional spectroscopic techniques, such as phase cycling schemes. Particularly helpful in this regard, is the ability to engineer different decay, disorder and dephasing sources in the Rydberg setting. We demonstrated this with several simple arrangements of a Rydberg dimer and Rydberg trimer embedded in a controllable environment, using realistic parameters.

To illustrate the points above, we did not need to vary the third time delay in our four pulse sequence, the so called waiting time $T$. By additionally scanning the waiting time, one could not only acquire information about system eigenstates, but also about time-evolution of the system in a non-equilibrium situation \cite{engel_fleming:coherence_nature}. In one further step, by Fourier transforming the third delay, the technique would be augmented to 3D spectroscopy \cite{Hayes_FMO_3Dspec}. Finally, the technique can in principle be augmented to an increasingly larger number of pulses, furnishing ND-spectroscopy.

 Another technique that would be interesting to explore in a Rydberg system replaces phase cycling used here by phase modulation, which has become popular recently in optical multidimensional spectroscopy \cite{li_phasemodul_PRA,Tekavec_phasemodul_JChemP,Perdomo_Ortiz_phasemodul_JPCB,Damtie_2dimaction_PhysRevA,BrMuSt15_23877_}. In that technique the phase within a single pulse is modulated, and signals of interest in the end are extracted using knowledge of the specific phase profile used. A thorough study of a prospective experimental implementation of this scheme in the Rydberg setting would be needed to assess whether it can also benefit from quantum simulations.

When going beyond the examples discussed above, using 2D spectroscopy on larger, more theoretically challenging Rydberg systems could also add a window on Rydberg dynamics that is complementary to the many existing techniques, providing new ways to look at dipolar-interacting quantum systems, especially those interacting with their environment \cite{aman_dressloss_PhysRevA}, involving many-body resonant interactions \cite{Ryabtsev_foerster_PhysRevA}, spin relaxation \cite{Orioli_spinrelax_PRL} or transport \cite{whitlock_diffusivetransp}.

While we have chosen parameters for demonstrations in this article to yield most tractable theory, performing experiments would of course be even more interesting under conditions that challenge theory, in particular when the environment-aggregate system becomes non-Markovian \cite{genkin:markovswitch,schoenleber:thermal}. This can also be achieved for example
by allowing motion of the Rydberg aggregate atoms \cite{cenap:motion,wuester:cradle,wuester:CI} to be induced by dipole-dipole interactions, which then can mimick for example molecular vibrations \cite{roden:nmqsd}.

\ack
We gratefully acknowledge fruitful discussions with Ghassan Abumwis, Tan Howe-Sian, Shaul Mukamel, Matthias Weidem{\"u}ller and Pan-Pan Zhang and financial support from the Max-Planck society under the MPG-IISER partner group program as well as the Indo-French Centre for the Promotion of Advanced Research - CEFIPRA. AE acknowledges support from the DFG via a Heisenberg fellowship (Grant No EI 872/5-1). SW acknowledges support from the Investissements d'Avenir programme through the Excellence Initiative of the University of Strasbourg (IdEx) and the University of Strasbourg Institute for Advanced Study (USIAS).

\appendix

\section{Model for Rydberg aggregate embedded in a decohering environment}
\label{app_LeffHeff}

The  model considered in the present work, shown in \fref{EIT_level_scheme}, is discussed in detail in \rref{schoenleber:immag}. 
However, in that work the focus was on the single excitation manifold, i.e., states of the form $\ket{s,\cdots,p,\cdots,s}$ with a single $p$-excitation. 
 For 2D spectroscopy one must include also the total ground state and the states of the two exciton manifold.
We thus provide an extension of the approximate effective master equation that includes those states in this appendix.
 
 The total Hamiltonian, without the microwave, can be divided into three parts
\begin{eqnarray}
	\hat{H}=\sub{\hat{H}}{int}+\sub{\hat{H}}{EIT}+\sub{\hat{H}}{agg}.
	\label{Htot_app}
\end{eqnarray}
Where $\sub{\hat{H}}{agg}$ is the Rydberg aggregate Hamiltonian \bref{Hagg}.
The Hamiltonian $\sub{\hat{H}}{EIT}$ of the environment atoms that are driven by two lasers reads
 \begin{eqnarray}
 \sub{\hat{H}}{EIT}=& \sum\limits_{\alpha} \frac{\Omega_p}{2} ([\ket{e}\bra{g}]_{\alpha}+\text{H.c.})+\frac{\Omega_c}{2} ([\ket{r}\bra{e}]_{\alpha}+\text{H.c.}) \nonumber\\
 &-\Delta_p[\ket{e}\bra{e}]_{\alpha} -(\Delta_p+\Delta_c)[\ket{r}\bra{r}]_{\alpha},
 \label{eqnEIT}
 \end{eqnarray}
 where $\ket{g}$, $\ket{e}$ and $\ket{r}$ are ground state, excited state and Rydberg state. All of these are involved in the EIT scheme for the environment atoms \cite{schoenleber:immag}, which are labelled with the index $\alpha$. $\Omega_p$ and $\Omega_c$ are probe and coupling Rabi frequencies and  $\Delta_p$ and $\Delta_c$ are the respective detunings.
The interaction between environment atoms and aggregate atoms is given by
\begin{eqnarray}
\sub{\hat{H}}{int}&=\sum\limits_{k }^{2^N}\sum\limits_{\alpha} \bar{V}_{k\alpha} [\ket{r}\bra{r}]_\alpha \otimes \ket{\chi_k}\bra{\chi_k},
\label{int_hamil}
\end{eqnarray}
The many-body states $\ket{\chi_k}$ are defined in Eq.~\eref{manybodystates} of the main text  and $\bar{V}_{k\alpha}$ reads,
\begin{eqnarray}
\bar{V}_{k\alpha}&=\sum\limits_{i}^{N} 
\left\{\begin{array}{lr}
        V_{rs}^{i\alpha}, & \text{for }T_i^k=s\\
        V_{rp}^{i\alpha}, & \text{for }T_i^k=p\\
        \end{array}\right.   .
\label{vander_interaction}
\end{eqnarray}
The interaction potential between the $i$-th aggregate atom in the state $s$ and the $\alpha$-th environment atom in the Rydberg state $\ket{r}$ is given by  $V_{rs}^{i\alpha}=\frac{C_{6,rs}}{| r_{\alpha}-r_k |^6}$. 
For the case when the aggregate atom is in the $p$ state we discuss two different types of potential which can be selected by the choice of $\ket{r}$, 
$V_{rp}^{i\alpha}=\frac{C_{4,rp}}{| r_{\alpha}-r_k |^4}$,
or
  $V_{rp}^{i\alpha}=\frac{C_{6,rp}}{| r_{\alpha}-r_k |^6}$, as discussed in \rref{schoenleber:immag}. These different choices are also indicated in \tref{parameter_table}.
 
 Finally one has to take into account the spontaneous decay of the intermediate state, on the probe transition from $\ket{e}$ back to $\ket{e}$, with rate $\Gamma_p$. 
  This is accomplished by augmenting the equation of motion to a Lindblad master equation with decay operators $\hat{L}_\alpha=[\ket{g}\bra{e}]_\alpha$, as usual \cite{drake:atomicphysics}.

\subsection{Approximate effective master equation}

The full many-body Hilbertspace with many environment atoms and several aggregate atoms is too large to handle numerically.
For a wide range of laser parameters and couplings it is possible
 to adiabatically eliminate all excited states $\ket{e}$ and $\ket{r}$ of the environment atoms to obtain an evolution equation in the space of aggregate atoms alone. 
 Following the method of \rref{sorensen:adiabelim} we define a projector, 
 $P_g=\sum\limits_{k } \ket{\chi_k}\bra{\chi_k}\otimes \ket{\boldsymbol{g}}\bra{\boldsymbol{g}}, $
  on the relevant state space, in which all environment atoms are in the ground state $\ket{g}$.
 The first part acts on aggregate atoms and second part on environment atoms ($\ket{\boldsymbol{g}}=\ket{gg...g}$).
  The complementary operator to $\hat{P}_g$ is given by $\hat{P}_e=1-\hat{P}_g$.
  As a result of the adiabatic elimination procedure \cite{sorensen:adiabelim}, we obtain an effective Hamiltonian and Lindblad operators, $\sub{\hat{H}}{eff}$ and $\sub{\hat{L}}{eff}^{(\alpha)}$, which can be written as
$
\sub{\hat{H}}{eff}=-\frac{1}{2}\hat{\Omega}_-[\sub{\hat{H}}{NH}^{-1}+(\sub{\hat{H}}{NH}^{-1})^{\dagger}]\hat{\Omega}_+ + \hat{H}_g
$
and
$
\sub{\hat{L}}{eff}^{(\alpha)}=\hat{L}\sub{\hat{H}}{NH}^{-1}\hat{\Omega}_+
$,
where $\sub{\hat{H}}{NH}=\hat{H}_e-i\sum_\alpha \hat{L}_{\alpha}^{\dagger}\hat{L}_{\alpha}/2$ is a non-Hermitian Hamiltonian, while $\hat{H}_g$, $\hat{H}_e$, $\hat{\Omega}_+$ and $\hat{\Omega}_-$ are given by
$
\hat{H}_g=\hat{P}_g \hat{H} \hat{P}_g=\hat{P}_g \sub{\hat{H}}{agg} \hat{P}_g,
$
and \\
$
	\hat{H}_e
			=\hat{P}_e \bigg[\sum\limits_{\alpha} \frac{\Omega_c}{2} ([\ket{r}\bra{e}]_{\alpha}+\text{H.c.}) 
			-\Delta_p[\ket{e}\bra{e}]_{\alpha}-(\Delta_p+\Delta_c)[\ket{r}\bra{r}]_{\alpha} + \sub{\hat{H}}{int} + \sub{\hat{H}}{agg}\bigg] \hat{P}_e
$,
\\
$
\hat{\Omega}_+=\hat{P}_e \bigg[\sum\limits_{\alpha} \frac{\Omega_p}{2} [\ket{e}\bra{g}]_{\alpha} \bigg]\hat{P}_g$ and $\hat{\Omega}_-=\hat{\Omega}_+^{\dagger}
$.
\\
Evaluating the expressions above, and using the definitions \bref{main_Heff} and \bref{main_Leff}, 
we obtain 
\begin{equation}
\sub{h}{eff}^{(k)}=\bigg[\sum\limits_{\alpha}\frac{(4\tilde{V}_{k\alpha}\Delta_p+1)\Omega_p^2\tilde{V}_{k\alpha}}{16\tilde{V}_{k\alpha}^2|\tilde{\Delta}_p|^2+8\tilde{V}_{k\alpha}\Delta_p+1} \bigg],
\end{equation}
and
\begin{eqnarray}
\sub{\gamma}{eff}^{(k\alpha)} = & - \frac{2\tilde{V}_{k\alpha} \Omega_p\sqrt{\Gamma_p}}{1+4\tilde{V}_{k\alpha}\tilde{\Delta}_p},
\label{Lalpha}
\end{eqnarray}
where $\tilde{V}_{k\alpha}=(\bar{V}_{k\alpha}-\Delta_p-\Delta_c)/\Omega_c^2$.  We have defined $\tilde{\Delta}_p=\Delta_p+i\Gamma_p/2$. As long as $\ket{\chi_k}$ contains only a single excitation, this reduces to the expressions in \cite{schoenleber:immag}.

Under the conditions discussed in \sref{environment}, we now can solve \bref{mastereqn} using the results of the present section to predict all measurements
on the Rydberg aggregate, including 2D spectra. This is computationally much easier than the full many-body evolution \bref{generic_evolution_equation}.

\section{Perturbative calculation of system response}
\label{app_perttheory}

To understand the spectra shown in the present work, it is useful to consider analytical expressions obtained from standard perturbation theory~\cite{mukamel:book} with respect to the aggregate-radiation interaction. For reference we provide here the basic equations needed in the Rydberg context.

We first diagonalize $\sub{\hat{H}}{agg}+\hat H_{\rm eff}$ to obtain the eigenenergies and eigenstates of the Rydberg aggregate system
\begin{equation}
\left(\sub{\hat{H}}{agg}+\hat H_{\rm eff}\right)\ket{k}= {\cal E}_k \ket{k}.
\label{effhamil_SE_appendixC}
\end{equation}
Both, $\hat H_{\rm agg}$ and $\hat H_{\rm eff}$, conserve the number of $p$-excitations, $N_p$, hence the eigenstates can be classified according to $N_p$. 
States with $N_p=1$ are denoted as one-exciton states and states with $N_p=2$ as two exciton states.

For zero microwave detuning, \eref{Hmw} becomes $\hat H_{mw}=E(t) \sum_{n=1}^N \hat\mu_n$ with $\hat{\mu}_n=\mu_0( \hat{\sigma}_\mathrm{sp}^{(n)} + \hat{\sigma}_\mathrm{ps}^{(n)})/2$.
The matrix elements of $\hat\mu\equiv \sum_n \hat{\mu}_n$ between eigenstates \bref{effhamil_SE_appendixC} are denoted by $\bra{\tilde\chi_k}\hat\mu\ket{\tilde\chi_{k'}}=\mu_{kk'}$ and determine whether or not transitions between these states are allowed. The time evolution in \eref{mastereqn}, including the Lindblad terms from the environment, is written as
\begin{eqnarray}
\dot{ \rho}(t) &=-\frac{i}{\hbar}\breve{\cal L}_0\rho -\frac{i}{\hbar}\breve{\cal L}_{mw}(t)\rho,
\end{eqnarray}
where $\breve\mathcal{L}_0$ is the super-operator, indicated by $\breve{ }$, accounting for free evolution of the system without the microwave field.

Following~\cite{mukamel:book}, we expand the time-evolving $\rho$ to fourth order in the electric field such that the relevant contribution to our signal is
\begin{eqnarray}
\label{4th order sig}
S(t) = \mathrm{Tr}\{ \hat{F} \rho^{(4)}\},
\end{eqnarray}
with $\hat F$ defined in \sref{ryd_aggregate}. After the expansion, we find
\begin{eqnarray}
S(t)&=\displaystyle\frac{1}{\hbar^4} \mathrm{Tr}\{ \hat{F}
\int_{t_0}^td\tau_4\int_{t_0}^{\tau_4}d\tau_3\int_{t_0}^{\tau_3}d\tau_2\int_{t_0}^{\tau_2}d\tau_1\breve\mathcal{G}(t-\tau_4)\breve{\cal L}_{mw}(\tau_4)
\nonumber\\
&\times\breve\mathcal{G}(\tau_4-\tau_3)\breve{\cal L}_{mw}(\tau_3)\breve\mathcal{G}(\tau_3-\tau_2)\breve{\cal L}_{mw}(\tau_2)
\nonumber\\
&\times\breve\mathcal{G}(\tau_2-\tau_1)\breve{\cal L}_{mw}(\tau_1)\breve\mathcal{G}(\tau_1-t_0)\rho(t_0)\},
\end{eqnarray}
with the Green's matrix in the time domain defined by
\begin{eqnarray}
\breve\mathcal{G}(\tau)&=\theta(\tau)\exp\bigg(-\frac{i}{\hbar}\breve{\cal L}_0\tau\bigg),
\label{propagator}
\end{eqnarray}
From \bref{Hmw} we see that we can write $\breve\mathcal{L}_{\rm mw}= \breve\mathcal{V} E(t)$, splitting the light-matter interaction into an operator and an electric field part, see \eref{Hint_generic}.
Substituting $t_1 = \tau_2-\tau_1,t_2=\tau_3-\tau_2,\ldots t_4 = t-\tau_4$ and making use of the initial time, $t_0\rightarrow-\infty$, we obtain, 
\begin{eqnarray}
\label{corr-mat}
S(t)
&=\displaystyle\frac{1}{\hbar^4}
\int_{0}^\infty dt_4\int_{0}^{\infty}dt_3\int_{0}^{\infty}dt_2\int_{0}^{\infty}dt_1\nonumber\\
&\times  {\cal R}(t_1,t_2,t_3,t_4){\cal E}(t,t_4,t_3,t_2,t_1)
\end{eqnarray}
with the response function
\begin{eqnarray}
\label{CCM}
 {\cal R}(t_1,t_2,t_3,t_4)&= \mathrm{Tr}\{ \hat{F}\breve\mathcal{G}(t_4)\breve\mathcal{V}\breve \mathcal{G}(t_3)\breve\mathcal{V}\breve\mathcal{G}(t_2)\breve\mathcal{V}\breve\mathcal{G}(t_1)\breve\mathcal{V}\rho(-\infty)\},
\end{eqnarray}
and electric field product
\begin{eqnarray}
{\cal E}(t,t_4,t_3,t_2,t_1)&= E(t-t_4)E(t-t_4-t_3)\nonumber\\
&\times E(t\!-\!t_4\!-\!t_3-t_2)E(t\!-\!t_4\!-\!t_3\!-\!t_2\!-\!t_1).
\end{eqnarray}
Let us finally decompose the atom-field coupling super-operator $\breve\mathcal{V}$ into $\breve{\mathcal{V}} =\breve{\mathcal{V}}_{\rm up}+ \breve{\mathcal{V}}_{\rm down}$ where the part $\breve{\mathcal{V}}_{\rm up}$ contains the elements $\mu \sum_n \hat{\sigma}_{ps}$ of \bref{Hmw} that cause system excitations, while $ \breve{\mathcal{V}}_{\rm down}$ contains those, $\mu \sum_n \hat{\sigma}_{sp}$, causing de-excitations. 
Hence $ \breve{\mathcal{V}}_{\rm down}$ = $\breve{\mathcal{V}}_{\rm up}^\dag$ 
In the original Master-equation \bref{mastereqn}, the operators in $\hat V$ will act on the density matrix from the left as well as from the right, since these actions originate from a commutator \cite{mukamel:book}. 
We can represent this by further splitting the super-operators defined above in a left- and right component  \cite{harbola2008superoperator,dorfman2016nonlinear}, according to 
$\breve{\mathcal{V}}_{\rm up} = \breve {\cal V}_L + \breve {\cal V}_{R} $ with 
\begin{eqnarray}
\breve {\cal V}_L^{}{\cal \rho} &\leftrightarrow  \hat V_{\rm up} \hat\rho,~ \mbox{and}~
 \breve {\cal V}_{R}^{}{\cal\rho}&\leftrightarrow \hat\rho\hat V_{\rm up}
\label{leftright_supops}
\end{eqnarray}
where $\hat{V}_{\rm up}$ is the respective operator in Hilbert space.
Equation \bref{CCM} is a sum of terms involving four interactions with the electro-magnetic field that involve excitations or de-excitations acting on the left or the right of the density matrix. Between these interactions, the system propagates either in a population or coherence of the density matrix according to \bref{propagator}.
Each of these terms can thus be represented by a double-sided Feynman diagram as given for example in \fref{2dspectra_dimer}.

At this stage, the fourth order response \bref{corr-mat} contains still contributions that are not of much interest, such as such as non-resonant terms or the square of second order responses.
These can be removed through phase-matching when the system probed provides a medium for the interrogating fields, or through phase cycling when that is not the case. 

\section{Phase cycling technique}
\label{app_phasecycling}

After obtaining the simple expression for the fourth order response of our system in \bref{corr-mat}, we now give a brief idea how to isolate contributions of interest using phase-cycling. More information is widely available, \cite{tan:2dcycling,Zhang_5thorder_3Delec_JPCB,Meyer2000,bodenhausen1984selection, bain1984coherence,ZhEi16_4488_,de2014two}. 

We now also perform the rotating wave approximation (RWA), to ultimately reach the form \bref{Eoft_complex} for the fields in the light-matter coupling Hamiltonian. Through the RWA, the complex phase $e^{i \phi_j}$ of the $j$'th pulse effectively enters $\breve {\cal V}_L$, while $\breve {\cal V}_L^\dagger$ contains $e^{-i \phi_j}$. 
In contrast one obtains a factor $e^{-i \phi_j}$ in $\breve {\cal V}_R$, while $\breve {\cal V}_R^\dagger$ contains $e^{i \phi_j}$. This can be summarized by the rule that
all left-pointing arrows in diagrams as in \fref{2dspectra_dimer} contribute the phase $-\phi_j$, while right-pointing arrows are contributing with  $+\phi_j$. 

Regardless of these details, it is evident from the preceding discussion and \eref{corr-mat} and \eref{4th order sig}, that the dependence of the signal on the phases of a given single pulse sequence will formally take the form
\begin{eqnarray}
S_{\phi_1\phi_2 \phi_3 \phi_4}(t)&= \sum_{\alpha\beta\gamma\delta} \tilde{S}_{\alpha\beta\gamma\delta}(t) e^{i(\alpha \phi_1 + \beta \phi_2  + \gamma \phi_3 + \delta \phi_4 )},
\label{signal_phases}
\end{eqnarray}
where the integers $\alpha$ to $\delta$ count how many excitations or de-excitations occurred through a pulse with a given phase. We now recognise \bref{signal_phases} as discrete Fourier representation of the signal $S$, hence the coefficient $\tilde{S}_{\alpha\beta\gamma\delta}$ for a specific choice of indices can be extracted using an inverse Fourier transform from the space of phases to that of coefficients. 

Consider now the rephasing diagrams listed in \fref{2dspectra_dimer}. If we inspect the arrow directions and take into account the phase-sign allocation discussed at the beginning of this section, we find that all of them obtain a phase contribution  $\phi_{\rm reph}=-\phi_1+\phi_2+\phi_3-\phi_4$. In terms of the Fourier decomposition \bref{signal_phases} this corresponds to indices $\alpha=-1,\beta=1,\gamma=1,\delta=-1$. Since these are also \emph{all} diagrams with this combination of phases, we can obtain the signal corresponding to the sum of all rephasing diagrams by the inverse discrete Fourier transform:
\begin{eqnarray}
\bar S_{pe} \equiv \tilde{S}_{\alpha\beta\gamma\delta}(t)&= \sum_{\phi_1\phi_2 \phi_3 \phi_4} S_{\phi_1\phi_2 \phi_3 \phi_4}(t) e^{-i(\alpha \phi_1 + \beta \phi_2  + \gamma \phi_3 + \delta \phi_4 )},
\label{photon_echo_IDFT}
\end{eqnarray}
with the index choice $\alpha=-1,\beta=1,\gamma=1,\delta=-1$.

The final question is on how many different phase data-points $S$ in \bref{signal_phases} need to be sampled. Since each set of phases will take up experimental time, it is typically desirable to minimize the number of sets. Such schemes are discussed for example in Refs.~\cite{tan:2dcycling,dorineORIG}.
 We used a scheme which employs $27$ different sets of phases.
\begin{figure}[t]
	\centering
	\includegraphics[width=0.5\columnwidth]{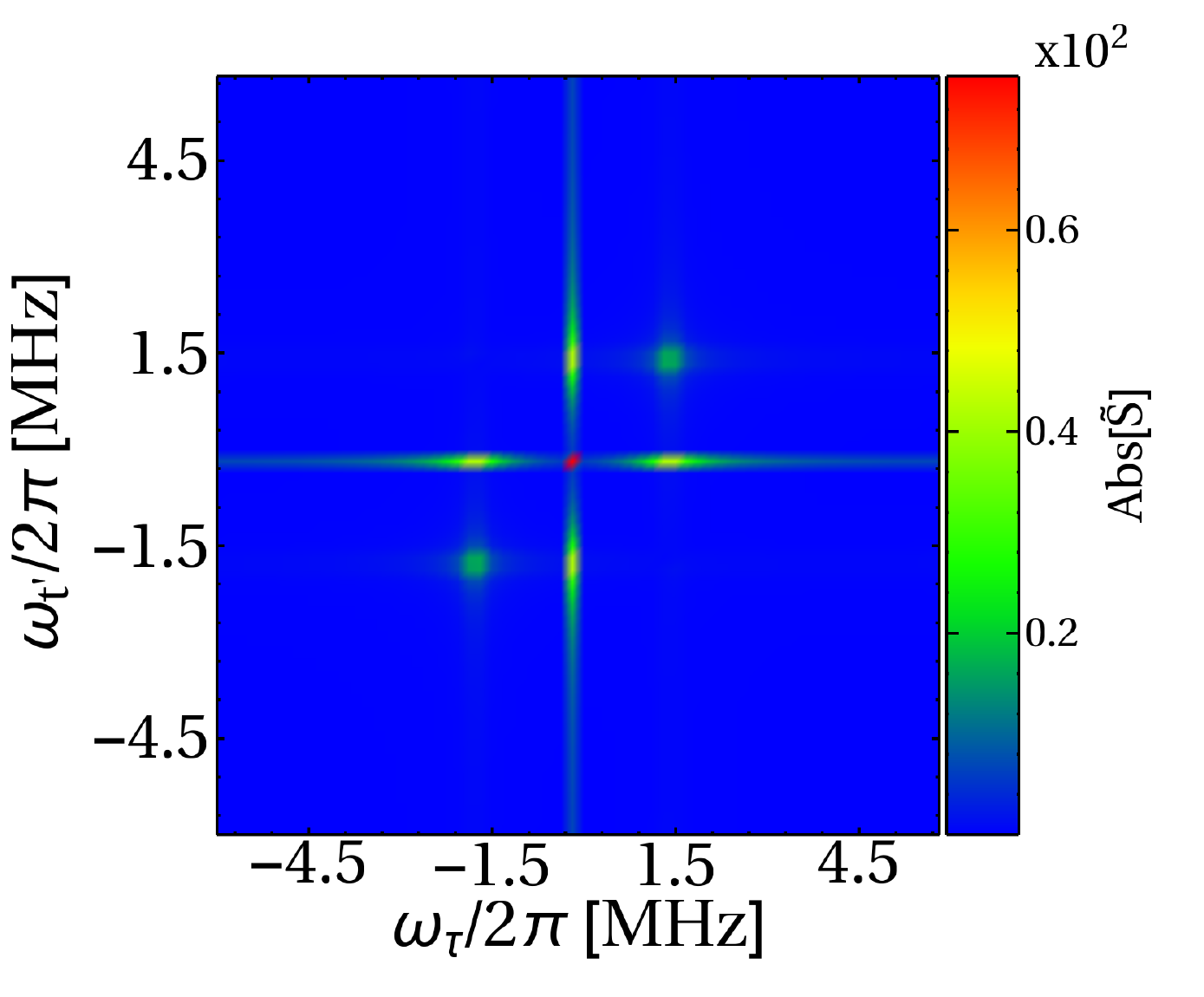}
	\caption{Two dimensional un-cycled spectrum of a monitored Rydberg dimer, for the same case as shown in \fref{2dspectra_dimer}(a). The signal intensity is clearly dominated by independent frequency features for each axis..
		\label{non_cycled}
	}
\end{figure}
Since only relative phases between pulses can matter, we set the phase of the last pulse to zero $\phi_4=0$, and then cycle the remaining ones in three discrete steps over the complex unit circle, $\phi_1=l (2\pi/3)$, $\phi_2=m (2\pi/3)$, $\phi_1=n (2\pi/3)$, with $l,m,n\in\{0,1,2\}$, yielding a total of $27$ combinations. Such a 27 step phase cycling scheme has been experimentally implemented in  two dimensional fluorescence spectroscopy \cite{de2014two}. 
Using this scheme, we thus explicitly obtain the phase cycled signal as
\begin{eqnarray}
\label{1pe-full}
\bar S_{pe}(\tau,T,t')&=\frac{1}{27}\sum_{l=0}^{2}\sum_{m=0}^{2}\sum_{n=0}^{2}
	e^{i(l\frac{2\pi}{3}-m\frac{2\pi}{3}
		-n\frac{2\pi}{3}})
 S_{(l\frac{2\pi}{3})(m\frac{2\pi}{3})(n\frac{2\pi}{3})} (\tau,T,t'),
\end{eqnarray}
where we have now made the remaining dependence on time delays explicit. Using
\begin{eqnarray}
\label{full_sig_num}
\tilde S_{pe}(\omega_\tau,T,\omega_{t'})=\frac{1}{2\pi}\displaystyle\int\int d\tau dt' \:\:e^{-i\omega_\tau\tau}
e^{i\omega_{t'}t'}\bar{S}_{pe}(\tau,T,t'),
\end{eqnarray}
this is then converted to frequency space ($\omega_\tau$, $\omega_{t'}$).

For the case $T=0$ one can reduce the number of phase-sets by using
\begin{eqnarray}
\label{1pe-full_18}
\bar S_{pe}(\tau,T=0,t')&=\frac{1}{27}\bigg[2\sum_{l=0}^{2}\sum_{m=0}^{2}\sum_{n=0}^{n<m}
e^{i(l\frac{2\pi}{3}-m\frac{2\pi}{3}
	-n\frac{2\pi}{3}})
S_{(l\frac{2\pi}{3})(m\frac{2\pi}{3})(n\frac{2\pi}{3})} (\tau,0,t') \nonumber \\
&+ \sum_{l=0}^{2}\sum_{m=0}^{2}
e^{i(l\frac{2\pi}{3}-m\frac{2\pi}{3}
	-m\frac{2\pi}{3}})
S_{(l\frac{2\pi}{3})(m\frac{2\pi}{3})(m\frac{2\pi}{3})} (\tau,0,t')\bigg],
\end{eqnarray}
which requires only 18 phase triples.

 In a similar way, schemes for the so-called non-rephasing signal ($\phi_{\rm non-reph}=\phi_1-\phi_2+\phi_3-\phi_4$) and the double quantum coherence signal ($\phi_{\rm doub}=\phi_1+\phi_2-\phi_3-\phi_4$) can be obtained, see diagrams in the supplemental material.

For a further intuitive idea on the functionality of phase cycling, consider the un-cycled spectrum in \fref{non_cycled}, obtained from simulations of \eref{mastereqn} for just a single set of phases: $\phi_1=\phi_2=\phi_3=\phi_4=0$. This corresponds to the Rydberg dimer of \fref{2dspectra_dimer}(a). Clearly, the peak structure of \fref{2dspectra_dimer}(a) is not visible. Instead the spectrum is dominated by second order features at $\omega_{\tau}=0$ (or $\omega_{t'}=0$), which imply a signal that is independent of the respective delay $\tau$ ($t'$). This in turn implies that this signal is independent of pulse one (three). Whenever a contribution to $S$ does not depend on one of the first three pulses, it is evident from 
\bref{1pe-full} that it drops out in the summation over complex pre-factors, since e.g.~$\sum_{n=0}^2 e^{in \left(\frac{2\pi}{3}\right)}=0$. In this manner leading second order contributions are removed and fourth order ones brought to the fore.

\section{Perturbative calculation of rephasing signal contributions}
\label{photon_echo}

Let us now continue where we left off in \aref{app_perttheory}, assuming that through phase-cycling as discussed in the previous appendix, the system response in \bref{CCM} is reduced to only those contributions corresponding to the  rephasing signals.
 We show only diagrams where the ket is excited, each diagram has a vertically mirrored partner, which contributes the complex conjugate amplitude, such that the final contribution to the population is real.
In the case of interaction on the ket, the first interaction term in (\ref{CCM}) is $\cal V_{R}$. To comply with the restrictions discussed in the previous \aref{app_phasecycling}, the rephasing signal component needs an equal number of inward and outward interactions, which reduces the number of terms in \eref{CCM} to four, for which the Feynman diagrams are shown in \fref{2dspectra_dimer}.
 Thus we obtain a response function ${\cal R}(t_1,t_2,t_3,t_4)= \mathcal{R}_1+\mathcal{R}_2+\mathcal{R}_3+\mathcal{R}_4$ with 
\begin{eqnarray}
\label{LSP_mat}
\mathcal{R}_1 &= -E_0^4  \mathrm{Tr}\{ \hat{F}\breve{\mathcal{G}}(t_4)\breve V_R^\dag \breve{ \cal{G}}(t_3)\breve V_L^{}\breve{\cal{G}}(t_2)\breve V_R^\dag\breve{ \cal{G}}(t_1)\breve V_R^{}\rho(-\infty)\},\\
\mathcal{R}_2 &=- E_0^4 \mathrm{Tr}\{ \hat{F}\breve{\cal{G}}(t_4)\breve V_R^{}\breve{\cal{G}}(t_3)\breve V_L^{\dag}\breve{\cal{G}}(t_2)\breve V_R^{}\breve{\cal{G}}(t_1)\breve V_R^{}\rho(-\infty)\},\\
\mathcal{R}_3 &= -E_0^4 \mathrm{Tr}\{ \hat{F}\breve{\mathcal{G}}(t_4)\breve V_L^{\dag}\breve {\cal{G}}(t_3)\breve V_L^{}\breve{\cal{G}}(t_2)\breve V_L^{}\breve{ \cal{G}}(t_1)\breve V_R^{}\rho(-\infty)\},\\
\mathcal{R}_4 &= + E_0^4 \mathrm{Tr}\{ \hat{F}\breve{\cal{G}}(t_4)\breve V_R^{}\breve{\cal{G}}(t_3)\breve V_L^{}\breve{\cal{G}}(t_2)\breve V_L^{}\breve{\cal{G}}(t_1)\breve V_R^{}\rho(-\infty)\}.
\label{LSP_mat_last}
\end{eqnarray}

The next step is to convolve the response function ${\cal R}$ with the electric field pulses as in \eref{corr-mat}, which is simplified by the impulsive limit where the electric field pulse shapes are delta functions, $A(t)=\delta(t)$ in \eref{Eoft}.
 We also express the response in 
the eigen-basis of the effective Hamiltonian, $\ket{\tilde{\chi}_k}$. Finally taking the Fourier transform of the signal, each response in \bref{LSP_mat} contributes a term
\begin{eqnarray}
\tilde{S}_i(\omega_{\tau},T,\omega_{t'})= E_0^4\frac{1}{\hbar^4}\displaystyle\int_0^\infty\int_0^\infty \mathcal{R}_i(\tau,T,t') e^{-i \omega_{\tau} \tau} e^{i \omega_{t'} t'} d\tau dt'.
\end{eqnarray}
Since there are no other time-arguments in the $\mathcal{R}_j$, in this impulsive limit for the field, Fourier transforms essentially act directly on the Green's functions.

For all cases considered in this work the Liouville propagator is diagonal in the chosen basis, which is a rather special case and requires the absence of relaxation processes that move population between different eigenstates.
However that allows us to obtain simple expression for the 2D spectrum, as $\tilde S(\omega_{\tau},T,\omega_{t'})=\sum_{i=1}^4\tilde{S}_i(\omega_{\tau},T,\omega_{t'})$ with contributions
\begin{eqnarray}
\label{eq:lsp-frew}
\tilde{S}_1^{} &=-E_0^4\sum_{i,j}|\mu_{s\beta_{j}}|^2|\mu_{s\beta_{i}}|^2
{\tilde{\cal{ G}}}_{\beta_js}(\omega_\tau)
{\cal{ G}}_{ss}(T){\tilde{\cal{ G}}}_{\beta_is}(\omega_{t'}),\\
\tilde{S}_2^{} &=-E_0^4\sum_{i,j}\mu_{s\beta_{j}}\mu_{\beta_{j}s}|\mu_{s\beta_{i}}|^2{\tilde{\cal{ G}}}_{\beta_js}(\omega_\tau)
{\cal{ G}}_{\beta_j\beta_i}(T){\tilde{\cal{ G}}}_{\beta_is}(\omega_{t'}),\\
\tilde{S}_3^{} &=-E_0^4\sum_{i,j,k}\mu_{s\beta_{j}}\mu_{\beta_{j}\zeta_{k}}\mu_{\zeta_{k}\beta_{i}}\mu_{\beta_{i}s}
{\tilde{\cal{ G}}}_{\zeta_k\beta_i}(\omega_\tau)
{\cal{ G}}_{\beta_j\beta_i}(T){\tilde{\cal{ G}}}_{\beta_is}(\omega_{t'}),\\
\tilde{S}_4^{} &=2E_0^4\sum_{i,j,k}\mu_{s\beta_{j}}\mu_{\beta_{j}\zeta_{k}}\mu_{\zeta_{k}\beta_{i}}\mu_{\beta_{i}s}
{\tilde{\cal{ G}}}_{\zeta_k\beta_i}(\omega_\tau)
{\cal{ G}}_{\beta_j\beta_i}(T) {\tilde{\cal{ G}}}_{\beta_is}(\omega_{t'}),
\end{eqnarray}
where the indices $i$, $j$ run over all single exciton eigenstates $\ket{\beta}$ and $k$ runs over all two-exciton states $\ket{\zeta}$. 
The factor 2 in the last equation comes from the fact that the final state in this path-way contains two p-excitations.
All expressions use the Green's function in the frequency domain, given by
\begin{eqnarray}
\label{eq:G(omega)}
\tilde{{\cal G}}_{kk'}(\omega)=\frac{1}{\omega-\omega_{kk'}-i\Gamma_{kk'}},
\end{eqnarray}
where $\omega_{kk'} = \omega_k-\omega_{k'}$ is the transition frequency between states $\ket{k'}$ and $\ket{k}$, and 
\begin{eqnarray}
\label{eq:Gammkk}
\Gamma_{kk'}&= \sum_\alpha \bigg[\bra{k}\sub{\hat{L}}{eff}^{(\alpha)} \ket{k} \bra{k'}\sub{\hat{L}}{eff}^{(\alpha)\dagger} \ket{k'} \nonumber \\
&- \frac{1}{2}( \bra{k}\sub{\hat{L}}{eff}^{(\alpha)\dagger}\sub{\hat{L}}{eff}^{(\alpha)} \ket{k} + \bra{k'}\sub{\hat{L}}{eff}^{(\alpha)\dagger}\sub{\hat{L}}{eff}^{(\alpha)} \ket{k'} ) \bigg]
\end{eqnarray}
 is the effective decay rate in this state. In general $\Gamma_{kk'}$ is complex and the imaginary part contributes a shift to the transition energy.

To compare with full numerical calculations, we calculate the total signal, $S=\sum_i^4S_i^{}$, and typically plot the modulus of the resultant complex function. It is thus instructive to consider the absolute value of $\tilde{{\cal G}}_{kk'}(\omega)$, which is a Lorentzian with peak at $\omega_{kk'}-\mathrm{Im}[\Gamma_{kk'}]$
and a width given by $\mathrm{Re}[\Gamma_{kk'}]$. 
To perform a discrete Fourier transform, we additionally multiply the signal with a Gaussian window function 
\begin{eqnarray}
\kappa(\tau,t') = \exp{\bigg(-\frac{\tau^2 + t'^2}{\sigma_G^2} \bigg)},
\label{gaussian_window}
\end{eqnarray}
prior to transform, where the width which is chosen as $\sigma_G=0.3\sub{\tau}{max}$, where $\sub{\tau}{max}$ depends on the largest value of $\tau$ and $t'$.

We can now apply the results obtained so far to the Rydberg dimer and trimer aggregates discussed in the main article.

\subsection{Dimer spectrum}
\label{feynman_diag}

For the symmetric dimer discussed in \sref{dimerspec} of the main text, there are only two non-zero transition dipole matrix elements
($\mu_{ss,+}$ and $\mu_{+,pp}$).
Then \eref{LSP_mat}-\eref{LSP_mat_last} from above become
\begin{eqnarray}
\label{dimer-lsp}
{\cal R}_1^{} &=-|\mu_{ss,+}|^4
{\cal G}_{ss,+}(\tau){\cal G}_{ss,ss}(T){\cal G}_{ss,+}(t'),\\
{\cal R}_2^{} &=-|\mu_{ss,+}|^4
{\cal G}_{ss,+}(\tau){\cal G}_{+,+}(T){\cal G}_{ss,+}(t'),\\
{\cal R}_3^{} &=-|\mu_{ss,+}|^2|\mu_{pp,+}|^2
{\cal G}_{ss,+}(\tau){\cal G}_{pp,pp}(T){\cal G}_{pp,+}(t'),\\
{\cal R}_4^{} &=2 |\mu_{ss,+}|^2|\mu_{pp,+}|^2
{\cal G}_{ss+}(\tau){\cal G}_{pp,pp}(T){\cal G}_{pp,+}(t').
\end{eqnarray}
 Each of the above responses represent an individual Liouville space pathway, for which diagrams are shown in \fref{2dspectra_dimer}. After Fourier transform, we can finally write the complete signal as
\begin{eqnarray}
\tilde{S}(\omega_{\tau},0,\omega_{t'})&= \frac{E_0^4}{2\pi} \Big({|\mu_{ss,+}|^2}
\tilde{{\cal G}}_{ss,+}(\omega_{\tau})\Big) \nonumber \\
& \times  \bigg(-2|\mu_{ss,+}|^2\tilde{{\cal G}}_{ss,+}(\omega_{t'}) + \big|\mu_{pp,+}\big|^2
\tilde{{\cal G}}_{pp,+}(\omega_{t'}) \bigg).
\label{dimer_signal_final}
\end{eqnarray}
The energy differences, from \eref{effhamil_SE_appendixC}, effective decay rates from \eref{eq:Gammkk} and transition dipoles of all transitions are listed in \tref{dimer_evals}.
For transition dipoles see the discussion in \sref{dimerspec}.
\begin{center}
	\begin{table}
		\begin{tabular}{| c| c| c| c| c |} 
			\hline
			 	& Transitions  & Transition frequencies& Transition dipole& $\Gamma_{kk'}/(2\pi)$  \\ 
				&between states & $\omega_{kk'}/(2\pi)$ [MHz]  & $\mu_{kk'}/(2\pi)$ [MHz]  &  [MHz]\\
			\hline
			1  & $\ket{ss}\rightarrow\ket{+}$ &  1.56   & 2.83& 0.22 - 0.04$i$ \\ 
			\hline
			2  & $\ket{ss}\rightarrow\ket{-}$ &  -1.68   & $0$ & 0.22 - 0.04$i$ \\ 
			\hline
			3  & $\ket{+}\rightarrow\ket{pp}$ &  -1.68   & 2.83& 0.21 + 0.04$i$ \\ 
			\hline
			4  & $\ket{-}\rightarrow\ket{pp}$ &  1.56   & $0$  & 0.21 + 0.04$i$ \\ 	
			\hline
		\end{tabular}
		\caption{ \label{dimer_evals} Parameters entering the analytical calculations \bref{dimer_signal_final} for the spectrum shown in \fref{2dspectra_dimer}, as discussed in the text.}
	\end{table}
\end{center}

\begin{figure}[t]
	\centering
	\includegraphics[width=0.99\columnwidth]{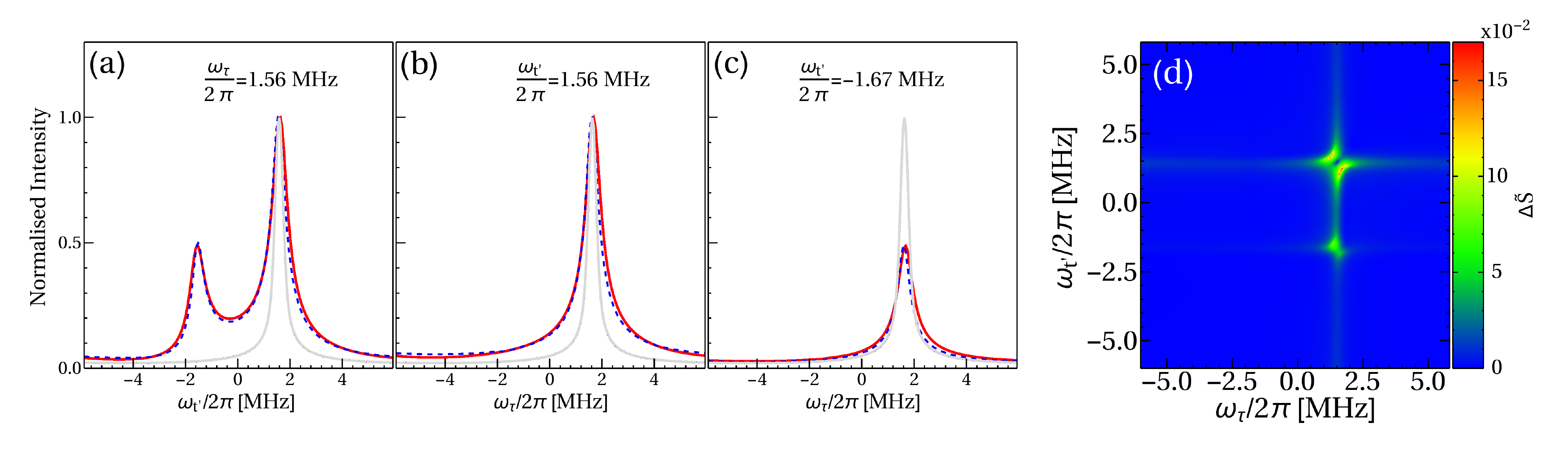}
	\caption{Comparison of results from numerical simulations of the Lindblad master equation, \eref{mastereqn}, and analytical calculations using four Liouville pathways shown diagrammatically in \fref{2dspectra_dimer}. (a) Cut along the $\omega_{t'}$ axis at $\omega_\tau/2\pi=1.56$ MHz, (red solid) numerical, (blue dashed) analytical and (light gray solid) Fourier transform of the half-sided Gaussian window function for orientation, shifted to the position of the main peak. (b) Cut along the $\omega_{\tau}$ axis at $\omega_{t'}/2\pi=1.56$ MHz and (c) cut along the $\omega_{\tau}$ axis at $\omega_{t'}/2\pi=-1.67$ MHz. (d) 2D spectrum of difference of the Fourier signals of numerical and analytical methods, $\Delta\tilde{S}=|\sub{\tilde{S}}{numerical}-\sub{\tilde{S}}{analytical}|$.
		\label{comparison_pic}	}
\end{figure}
It is instructive to compare the analytical result with the numerical non-perturbative calculations, based on the numerical solution of the complete master equation \bref{mastereqn}.
The results are shown in \fref{comparison_pic} along two selected one-dimensional cuts of \fref{2dspectra_dimer}. We find good agreement, validating both approaches. Residual small differences can be due to finite pulse duration or non-perturbative effects, which are included in the numerical solution, but not in the analytical one.
Note that the impulsive limit is not necessary to obtain analytical results, but it yields simpler expressions, which for our short pulses allow a direct interpretation of the 2D spectra.

\subsection{Trimer spectrum}
\label{trimer}

The eigen-states of the trimer system, with geometry shown in the inset of \fref{2dspectra_trimer}(b), are $\ket{sss},\ket{\beta_n},\ket{\zeta_n}$ and $\ket{ppp}$, where the $\ket{\beta_n}$ are given by
\begin{eqnarray}
\ket{\beta_1} &= 0.496 \ket{pss} - 0.711 \ket{sps} - 0.496 \ket{ssp} ,\nonumber 
\\ 
\ket{\beta_2}&= -0.503 \ket{pss} - 0.702 \ket{sps} + 0.503 \ket{ssp},\nonumber 
\\
 \ket{\beta_3}&= 0.707 \ket{pss}  + 0.707 \ket{ssp},\nonumber 
\end{eqnarray}
and the $\ket{\zeta_n}$ can be obtained from the $\ket{\beta_n}$ by swapping all $s$ states with $p$ states. As for the dimer, the interpretation of spectra in the main article will require all transition energies $\omega_{kk'}$ with the respective transition dipoles as well as  $\Gamma_{kk'}$ as provided in \tref{transition_table_trimer}.

\begin{center}
	\begin{table}
		\begin{tabular}{| c| c| c| c| c| } 
			\hline
			 & Transitions  & Transition frequencies& Transition dipole& $\Gamma_{kk'}/(2\pi)$ \\ 
				&between states & $\omega_{kk'}/(2\pi)$ [MHz]  & $\mu_{kk'}/(2\pi)$ [MHz]  &  [MHz]\\
			\hline
			1  & $\ket{sss}\rightarrow\ket{\beta_1}$ &  $-2.29$   & -1.42 & $0.11+0.04i$ \\ 
			\hline 
			2  & $\ket{sss}\rightarrow\ket{\beta_2}$ &  $2.29$ & -1.40 & $0.11+0.04i$\\ 
			\hline 
			3 & $\ket{sss}\rightarrow\ket{\beta_3}$  &  $-0.09$ & 2.83 & $0.12+0.04i $\\
			\hline 
			4 & $\ket{\beta_1}\rightarrow\ket{\zeta_1}$ & $-0.03$ & -0.99 &$ 0.24+0.06i$\\ 
			\hline 
			5  & $\ket{\beta_1}\rightarrow\ket{\zeta_2}$ & $ 4.55$ & 1.00 & $0.19+0.04i$\\ 
			\hline 
			6  & $\ket{\beta_1}\rightarrow\ket{\zeta_3}$ &  $2.17 $& -2.01 & $0.22+0.03i$ \\ 
			\hline 
			7 & $\ket{\beta_2}\rightarrow\ket{\zeta_1}$  &  $-4.61 $& 1.00 & $0.24+0.06i $\\
			\hline 
			8 & $\ket{\beta_2}\rightarrow\ket{\zeta_2}$ &  $-0.03 $& -1.01 & $0.19+0.04i$\\ 
			\hline 
			9  & $\ket{\beta_2}\rightarrow\ket{\zeta_3}$ & $-2.41 $& -1.99 & $0.22+0.03i$\\ 
			\hline
			10 & $\ket{\beta_3}\rightarrow\ket{\zeta_1}$  & $ -2.23$ & -2.01 &$ 0.22+0.06i$\\
			\hline 
			11 & $\ket{\beta_3}\rightarrow\ket{\zeta_2}$ &  $2.35 $& -1.99 & $0.19+0.04i$\\ 
			\hline 
			12  & $\ket{\beta_3}\rightarrow\ket{\zeta_3}$ & $-0.03 $& 2.00& $0.29+0.05i$ \\ 
			\hline 
		\end{tabular}
		\caption{ \label{transition_table_trimer} Parameters entering the analytic calculations for the Rydberg trimer (\fref{2dspectra_trimer}). }
	\end{table}
\end{center}
%

\section{Statistical Ensemble calculation}
\label{ensemble_calc}
In order to calculate the number of repetitions of the experiment required, we numerically determine the signal $S$ based on the count of a discrete number of Rydberg atoms in the $p$-state, 
as would be the case in experiment, rather than complete knowledge of the time evolved system density matrix $\hat{\rho}(t)$.

For this we first determine the reduced density matrix for a single aggregate atom, e.g.~for atom 1 in a dimer:
\begin{eqnarray}
	\hat{\rho}_1=\Tr_2\{\hat{\rho}\}=(\rho_{ss,ss}+\rho_{sp,sp})\ket{s}\bra{s}+(\rho_{ps,ps}+\rho_{pp,pp})\ket{p}\bra{p}.
\end{eqnarray}
From the resultant probabilities, a random number decides whether the atom is measured in $s$ or in $p$. This is repeated $\sub{N}{rep}$ times, to obtain a discretely sampled signal $\tilde{S}^{(\xi)}$ for each set of phases, which after phase cycling and Fourier transform gives rise to the graphs shown in \fref{fig:Nrep}. We conclude from this simulation, that about $\sub{N}{rep}\sim10^5$ repetitions are required to resolve the spectrum for these parameters.

\section{Trap center fluctuations}
\label{app_disorder_versus_cycling}

%
\begin{figure}[t]
	\centering
	\includegraphics[width=0.5\columnwidth]{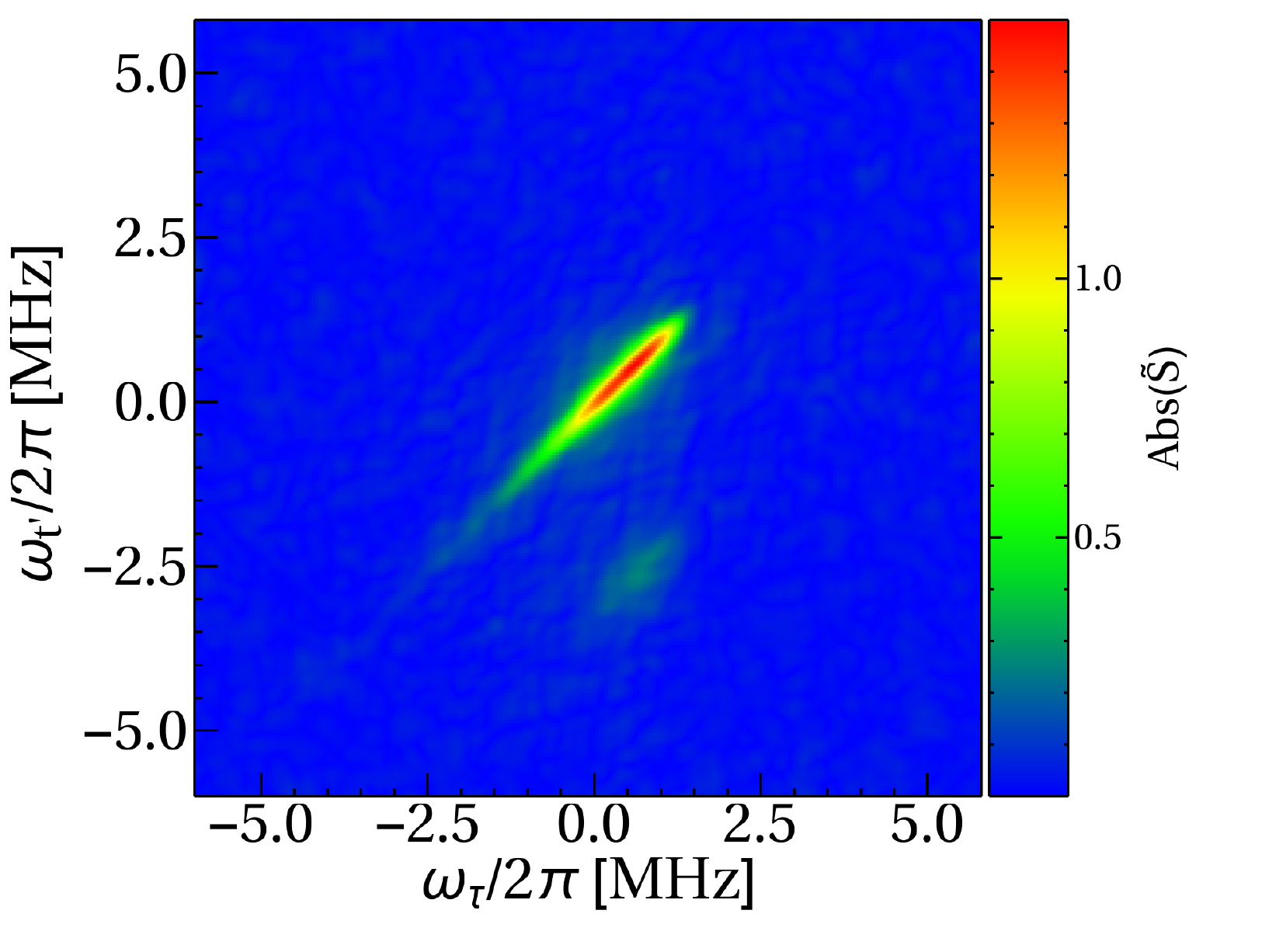}
	\caption{ Two dimensional spectrum of Rydberg dimer with position fluctuations during spectrum acquisition.
		Parameters are as in \fref{2dspectra_dimer}, except $\Omega_{mw}/(2\pi) = 16$ MHz, but the position of aggregate and detector atoms is given a random 2D offset with a Gaussian distribution of width $\sigma=200$ nm.
		This offset is varied after each individual set of four phases has been simulated. We use the technique of \aref{ensemble_calc} to model discrete atoms counts and average over $\sub{N}{rep}\sim10^5$ repetitions. \label{trap_agg_det}
	}
\end{figure}

In this appendix we explore the sensitivity of the phase-cycling procedure, discussed in \aref{app_phasecycling}, to position fluctuations of the aggregate atoms or the detector atoms.
Since we assume these to be trapped in the quantum ground state of their respective optical potential, such fluctuations would have to be due to imprecision in the alignment of these.

For the simulation, the position of each atom is given a random offset, drawn from a two-dimensional Gaussian distribution with standard-deviation $\sigma$. The random offset is newly drawn \emph{after each set of phases}, in contrast to the typical disorder situation in an optical setting. The results are shown in \fref{trap_agg_det} for $\sigma=200$ nm. 
The simulation employed the technique of the preceding appendix, and modelled the discrete counting of the number of $p$-excitatons in an experiment, averaging over $\sub{N}{rep}=10^5$ 
repetitions. We find, that for larger width $\sigma$ than shown, the small side-peak is no longer resolved so that $\sub{N}{rep}$ would have to be further increased.

\vspace{1cm}
\bibliography{TwoDimSpectroscopy_v4,refs_ALex}
\bibliographystyle{sebastian_v3}
\end{document}